\documentclass{ametsoc}
\usepackage[english]{babel}
\usepackage[T1]{fontenc}
\usepackage[latin9]{inputenc}
\usepackage{csquotes}
\usepackage{amsmath,amssymb}
\usepackage{textcomp} 
\usepackage{graphicx}
\graphicspath{{figures/}}
\renewcommand{\vec}[1]{\mathbf{#1}}

\journal{jas}
\abstract{Strong eastward jets at the equator have been observed in many planetary atmospheres and simulated in numerical models of varying complexity.
  However, the nature of the transition from a conventional state of the general circulation, with easterlies or weak westerlies in the tropics, to such a superrotating state remains unclear.
  Is it abrupt or continuous? This question may have far-reaching consequences, as it may provide a mechanism for abrupt climate change in a planetary atmosphere, both through the loss of stability of the conventional circulation and through potential noise-induced transitions in the bistability range.
  We study two feedbacks which may lead to bistability between a conventional and a superrotating state: the Hadley cell feedback and a wave-jet resonance feedback.
  We delineate the regime of applicability of these two mechanisms in a simple model of zonal acceleration budget at the equator.
  Then, we show using numerical simulations of the axisymmetric primitive equations that the wave-jet resonance feedback indeed leads to robust bistability, while the bistability governed by the Hadley cell feedback, although observed in our numerical simulations, is much more fragile in a multilevel model.}

\begin{document}

\title{Atmospheric bistability and abrupt transitions to superrotation:\\ wave-jet resonance and Hadley cell feedbacks}

\authors{Corentin Herbert\correspondingauthor{}}
\email{corentin.herbert@ens-lyon.fr}
\affiliation{Univ Lyon, ENS de Lyon, Univ Claude Bernard, CNRS, Laboratoire de Physique, F-69342 Lyon, France}
\extraauthor{Rodrigo Caballero}
\extraaffil{Department of Meteorology, Stockholm University, Stockholm, Sweden}
\extraauthor{Freddy Bouchet}
\extraaffil{Univ Lyon, ENS de Lyon, Univ Claude Bernard, CNRS, Laboratoire de Physique, F-69342 Lyon, France}

\maketitle

\section{Introduction}

A long standing question in the study of the general circulation of the atmosphere is the uniqueness of the solution, for fixed boundary conditions.
This problem was formulated early on by~\citet{Lorenz1967}, under the name of \emph{transitivity}.
This question is an important one, because it may have deep consequences on climate dynamics.
Indeed, in the presence of multiple attractors, the system may exhibit abrupt transitions from one to the other, induced either by internal variability or by an external forcing.
Paleoclimatic records provide evidence for such abrupt climate changes (e.g.\ Dansgaard-Oeschger events~\citep{Dansgaard1993}).
This type of events have so far been linked to nonlinear behavior of the oceanic circulation, whose bistability is fairly well documented~\citep{Dijkstra2005}, or of the coupling between several climate components~\citep{Boers2018}.
The atmosphere itself may admit multiple equilibria.
As a matter of fact, turbulent flows often exhibit coexisting steady-states for given external parameters, as well as spontaneous transitions between the two stable states, as has been reported in both numerical studies~\citep{Bouchet2009,Cortet2010,Bouchet2019} and laboratory experiments~\citep{Berhanu2007, Cortet2010, SaintMichel2013, Michel2016}.
Some of these experiments~\citep{Weeks1997, Tian2001} are actually inspired by geophysical flows~\citep{Charney1979}.
However, the question remains if such phenomena could occur at the level of the general circulation of the atmosphere.

An interesting candidate for bistability of the general circulation of the atmosphere is \emph{superrotation}~\citep{Held1999}: it refers to an atmospheric flow for which there exists a region carrying a larger angular momentum than the one associated to solid body rotation at the equator.
While the \emph{conventional} circulation of the atmosphere of the Earth has
mid-latitude westerly jets and weak easterlies in the tropics~\citep{Lee1999,Dima2005}
(and everywhere smaller angular momentum than the surface at the equator), a
superrotating atmosphere exhibits strong westerlies in the tropics.
This is actually observed on other planets of the Solar System~\citep{Read2018}, such as Jupiter~\citep{Porco2003}, Saturn~\citep{DChoi2009} (and its moon Titan~\citep{Flasar1998b}) or Venus~\citep{Marov1973}.
On Earth, superrotation may have played a role in the climate of the past: it was observed in numerical simulations of warm climates such as the Eocene~\citep{Caballero2010}, and it has been suggested that it could explain the permanent El Ni\~no conditions indicated by paleoclimatic proxies during the Pliocene~\citep{Tziperman2009}.
Another indicator of the robustness of superrotation is that it has been observed in numerical experiments with models of varying complexity: shallow-water models~\citep{Scott2008,Showman2010,Showman2011,Suhas2017}, two-level primitive equations~\citep{Suarez1992,Saravanan1993}, and multilevel comprehensive GCMs~\citep{Kraucunas2005,Schneider2009,Caballero2010,Showman2011,Arnold2012,Potter2014}.

A natural question to ask first is how is superrotation maintained at a dynamical level?
It is often stated that because of Hide's theorem~\citep{Hide1969}, superrotation must involve upgradient fluxes of angular momentum, which can only be achieved by eddy fluxes.
Let us give an alternative formulation of this argument: let us consider an axisymmetric model of the atmospheric circulation (for instance the axisymmetric primitive equations).
As can be easily checked, in the absence of forcing and dissipation, angular momentum is conserved by axisymmetric dynamics: $\partial_t M + \vec{u} \cdot \nabla M = 0$, with $M(\phi, p)=a \cos \phi(\Omega a \cos \phi + u(\phi, p))$, where $a$ and $\Omega$ are, respectively, the planetary radius and rotation rate, $\phi$ and $p$ the latitude and pressure coordinates, and $u$ the zonal wind.
Akin to vorticity conservation in 2D flows, this implies that all the norms of the angular momentum field are conserved, and in particular, its maximum.
This means that axisymmetric dynamics cannot evolve a sub-rotating atmosphere ($\max_{\phi, p} M(\phi, p) < \Omega a^2$) into a superrotating atmosphere ($\max_{\phi, p} M(\phi, p) > \Omega a^2$), and vice-versa.
There are potentially many ways eddies could accelerate the zonal wind towards the east in the tropics, and several routes to superrotation have already been found.
They can be broadly grouped into a series of studies which modify the basic physical
parameters of the planet, such as the planetary rotation rate~\citep{DiasPinto2014} or
the radius of the planet~\citep{Mitchell2010, Potter2014}.
The emerging scenario in this type of setup is that a hydrodynamic instability known as the \emph{Kelvin-Rossby instability}~\citep{Iga2005, PWang2014, ZuritaGotor2018}, generates the eddies which converge momentum in the tropics.
Instead of relying on an instability, a second thread of works has explored the possibility of stimulating wave emission from the tropics to account for equatorial momentum convergence, akin to the classical picture for mid-latitude jets~\citep{VallisBook}.
Enhanced wave activity in the tropics can be the result of several physical processes: convection, day-night contrast in tidally locked exoplanets~\citep{Merlis2010,Showman2011}, etc.
Broadly speaking, such processes can be modelled as non-zonal heating of the tropics: idealized GCM studies including such an additional forcing term have led to abrupt transitions to superrotation once the forcing amplitude reaches a certain threshold~\citep{Suarez1992, Saravanan1993, Kraucunas2005, Arnold2012}.

In fact, coexistence of the superrotating state with the conventional circulation for some range of parameters requires more than just eddy momentum flux convergence onto the equator.
Indeed, some \emph{positive feedback} mechanism is needed, so that the zonal-mean zonal wind budget may admit several solutions.
Such a feedback mechanism may come directly from the eddy forcing, or alternatively, from the mean meridional circulation.
The first possibility has been explored in particular by~\citet{Arnold2012}, who suggested a resonant feedback mechanism based on the properties of equatorial Rossby waves on a background mean-flow.
Relying on an explicit computation of the linear response of a shallow-water atmosphere to non-zonal tropical heating, in the spirit of the pioneering work of~\citet{Matsuno1966} and~\citet{Gill1980}, they have argued that the amplitude of the response depends on the background zonal wind in such a way that a resonance appears close to the opposite of the phase velocity of free Rossby waves.

The second possibility was suggested by~\citet{Shell2004} who showed that the Hadley cell itself could admit multiple equilibrium states.
Indeed, a conventional Hadley cell with updraft on the equator advects low momentum air into the upper troposphere, thereby inhibiting the onset of superrotation.
However, the contribution to the zonal momentum budget is the product of two terms: $\omega \partial_p u$.
These two terms behave differently: when westerly winds increase in the tropical upper troposphere, vertical shear increases, but vertical velocity decreases.
The net effect is a negative feedback in a first regime, then a positive feedback in a second regime.
Analytical arguments relying on a simplification of the zonal momentum budget at the equator, assuming that the Hadley cell, the imposed forcing and some frictional dissipation balance each other, show that this feedback leads to multiple equilibria.
These arguments are supported by numerical simulations in a very simple framework (1D axisymmetric shallow-water equations with a constant torque imposed).
A natural question to ask is whether this behavior remains in more realistic conditions.

In this paper, we explore the robustness of the two bistability mechanisms mentioned above: Hadley cell feedback and resonant response to equatorial heating.
First, we explicitly show in an analytical model how the resonant structure of the eddy momentum flux convergence can lead to bistability, and observe the corresponding hysteresis phenomenon in numerical simulations of the axisymmetric primitive equations.
Then, we ask whether the results of~\citet{Shell2004} extend to a multilevel model.
We find numerically that bistability may be obtained in this framework, but that it is relatively fragile.
Finally, we investigate the interplay between the two mechanisms.
We show that depending on the parameter characterizing the width of the resonant response to equatorial heating, two types of superrotating states can be found.
For wide resonances, the superrotating state has a weaker mean meridional circulation than the conventional state, and the range of forcing amplitudes for which both states coexist is quite small (Hadley cell-driven superrotation).
On the other hand, for narrow resonances, the strength of the mean meridional circulation does not change much across the bifurcation point, and the coexistence range is much wider (resonance-driven superrotation).

To reach these conclusions, we combine theoretical arguments obtained in a simplified framework based on the shallow-water model (Sec.~\ref{sec:shell}), and numerical simulations of the axisymmetric primitive equations (Sec.~\ref{sec:heldhou}).
A first obstacle to analytical progress comes from the fact that there is in general no closed form for the evolution equation of the zonal mean zonal wind.
Here, we take avantage of the fact that a linear response computation of the Matsuno-Gill type, for a constant and uniform zonal-mean wind $U$, has been found to agree relatively well with GCM results~\citep{Arnold2012}.
In Sec.\ref{sec:shell}\ref{sec:matsuno}, after recalling this computation, we describe the wave-jet resonance mechanism which provides a positive feedback.
Then, we present analytical arguments to disentangle the effects of the two nonlinear mechanisms: the wave-jet resonance and the Hadley cell, by studing the fixed-points of the zonal momentum budget at the equator (Secs.~\ref{sec:shell}\ref{sec:bistabilitybalancequalitative} and~\ref{sec:shell}\ref{sec:bistabilitybalancequantitative}), for different values of the forcing parameters.
Finally, we test the scenarii outlined through the analytical study of the shallow-water model in a more realistic model of the atmosphere, by carrying out numerical simulations of the 2D axisymetric primitive equations (Sec.~\ref{sec:heldhou}).

\section{Bistability in an analytical model of equatorial momentum balance}\label{sec:shell}

\subsection{The shallow-water model}

We first consider the simplest possible model which can account for both feedback mechanisms: a thin layer of fluid, described by the shallow-water equations, exchanging mass and momentum with a quiescent underlying layer.
The fluid is forced by diabatic heating $Q$ and dissipates energy through a Rayleigh friction $\epsilon$.
In spherical coordinates, these equations may be written as
\begin{align}
  \partial_t u + \frac{u}{a \cos \phi} \partial_\lambda u + \frac{v}{a\cos \phi} \partial_\phi (u \cos \phi) -2\Omega \sin \phi v
  &= - \frac{g}{a \cos \phi}\partial_\lambda h - \epsilon u +R_u, \label{eq:sweusphere}\\
  \partial_t v + \frac{u}{a \cos \phi}\partial_\lambda v + \frac{v}{a} \partial_\phi v + \frac{u^2}{a} \tan \phi + 2\Omega \sin \phi u
  &= -\frac{g}{a} \partial_\phi h -\epsilon v +R_v, \label{eq:swevsphere}\\
\partial_t h + \frac{1}{a \cos \phi}\lbrack \partial_\lambda (hu) +\partial_\phi(hv\cos\phi) \rbrack &= Q, \label{eq:swehsphere}
\end{align}
where $u$ and $v$ are the zonal and meridional components of the wind, $h$ the thickness of the fluid layer, $\lambda$ the longitude coordinate, $g$ the acceleration of gravity, $\Omega$ the rotation rate and $a$ the planetary radius.
The mass source/sink term $Q$ accounts both for radiative forcing and an additional non-zonal heating term, confined in the tropics, which represents in a rough manner convective effects or day-night contrast in tidally locked exoplanets.
As a consequence, the fluid layer also exchange momentum with the underlying \enquote{sponge} layer through the terms $R_u=-Qu/h \Theta(Q)$ and $R_v=-Qv/h \Theta(Q)$, where $\Theta$ is the Heaviside function.
This mechanism provides a rudimentary representation of the Hadley cell in the shallow-water model.

We now decompose all the fields into their zonal average, denoted by an overbar, and their eddy component, denoted by a prime: $u = \bar{u}+u'$, $v=\bar{v}+v'$, $h=\bar{h}+h'$.
In this context, the zonal mean wind profile $\bar{u}(\phi)$ satisfies the equation:
\begin{equation}
  \partial_t \bar{u} + \frac{\bar{v}}{a\cos \phi} \partial_\phi (\bar{u}\cos\phi)-2\Omega \sin \phi \bar{v} = -\frac{1}{a\cos\phi}\overline{v'\partial_\phi(u'\cos \phi)}+ \bar{R}_u - \epsilon \bar{u}. \label{eq:swezonalwind}
\end{equation}
Our goal is to study the possibility of multiple equilibria in this equation.
In general, this depends on the form of the eddy momentum flux convergence $F=-\frac{1}{a\cos\phi}\overline{v'\partial_\phi(u'\cos \phi)}$.
In a first step, we assume that it does not depend on $\bar{u}$ and discuss the other feedback mechanism $\bar{R}_u$, associated to the Hadley cell (Sec.~\ref{sec:shell}\ref{sec:bistabilityhadley-balance}).
We shall discuss the wave-mean flow interaction in Sec.~\ref{sec:shell}\ref{sec:matsuno}.

\subsection{Simplified zonal momentum balance at the equator}\label{sec:bistabilityhadley-balance}

At the equator, in a perpetual equinox configuration, the steady-state zonal-mean zonal momentum budget~\eqref{eq:swezonalwind} reduces to a balance between the eddy forcing $F$, the vertical advection by the Hadley cell $\bar{R}_u$ and the frictional dissipation.
This balance writes $F+\bar{R}_u-\epsilon \bar{u}=0$.
In this section, we study the existence of multiple solutions to this balance equation.
Reducing this way the problem to a zero-dimensional model allows for a qualitative understanding of the physical mechanisms potentially leading to multiple equilibria, as it is easy in this case to draw the different terms of this balance relation as functions of the parameters of the problem.
As a matter of fact, \citet{Shell2004} have shown that, for a constant forcing $F$, this simple zonal momentum balance model demonstrates how the Hadley cell feedback can lead to bistability.
We recall their argument in this section.
As we shall only be working with zonally averaged fields, we should drop the overbar from now on.
The value of all fields at the equator will be denoted with a null subscript.

Modelling radiative forcing by a Newtonian relaxation $Q=(h_{\text{eq}}-h)/\tau$ to a prescribed radiative equilibrium profile $h_{\text{eq}}$ with relaxation time $\tau$, and recalling that $R=-Qu/h \Theta(Q)$, the balance relation becomes
\begin{equation}\label{eq:shellubalance}
  F_0 -\epsilon u_0 + \frac{u_0}{h_0}\frac{h_0-h_{0\text{eq}}}{\tau} = 0, \text{ for } h_0 < h_{0\text{eq}}.
\end{equation}
A relation between the layer thickness and the zonal wind velocity can be obtained through a simple model of the Hadley cell~\citep{Held1980, Shell2004, VallisBook}.
The idea is that the thickness $h$ is in geostrophic equilibrium with the angular momentum conserving wind $u_m = \frac{u_0+\Omega a \sin^2\phi}{\cos \phi}$ in the tropics:
$(g^*/a) \partial_\phi h = -2\Omega u_m \sin \phi$, and in radiative equilibrium: $h=h_{\text{eq}}$ outside.
Integrating the geostrophic equilibrium equation and matching the resulting profile with the radiative equilibrium at a latitude determined by mass conservation yields
\begin{equation}\label{eq:shellhbalance}
  h_0-h_{0\text{eq}}=-\frac{5}{18 g^*} {(u_{0\text{eq}}-u_0)}^2.
\end{equation}
Note that the model only makes sense for $u_0 < u_{0\text{eq}}$.
Introducing the non-dimensional variables $U$ and $H$ through $h_0 = H h_{0\text{eq}}$, $u_0 = U u_{0\text{eq}}$, the two equations~\eqref{eq:shellhbalance} and~\eqref{eq:shellubalance} reduce to the simple algebraic system
\begin{align}
  1-H  &= p(U-1)^2,\\
  q  &= (1-H)U+rU,
\end{align}
where we have assumed $H\approx 1$, with
\begin{equation}
  p=\frac{5u_{0\text{eq}}^2}{18 g^\star h_{0\text{eq}}}, \quad q=\frac{F\tau}{u_{0\text{eq}}}, \quad r=\epsilon\tau.\label{eq:shellparams}
\end{equation}
Parameter values are given in table~\ref{tab:shellheld}.
\begin{table}
\begin{tabular}{ccccccc}
  $u_{0\text{eq}}$ & $h_{0\text{eq}}$ & $g^\star$ & $\tau$ & $\epsilon$ & $p$ & $r$ \\
  \hline
  60 m.s\textsuperscript{-1} & 16500 m & 0.08 $g$ & 8.10\textsuperscript{5} s & 10\textsuperscript{-8} s\textsuperscript{-1} & 0.077 & 0.008
\end{tabular}
\caption{Parameter values from the~\citet{Shell2004} model.}\label{tab:shellheld}
\end{table}
Hence, the balance between the forcing, Rayleigh friction, and the Hadley cell advecting low momentum wind from the lower layer is governed by the equation
\begin{equation}\label{eq:simplemomentumbalance}
  pU{(U-1)}^2+rU-q=0.
\end{equation}
This theory makes the feedback structure of the Hadley cell very clear: it acts as a positive feedback between the two roots of its derivative, $1/3 < U < 1$, and as a negative feedback for $U < 1/3$ and for $U > 1$.
When the forcing term is a constant imposed torque, the equation is a simple cubic equation, and the condition for bistability can be easily obtained.
A necessary condition is $0 \leq r/p \leq 1/3$: it is the condition for the function $pU{(U-1)}^2+rU$ to have a local maximum.
With the default parameter values, $r/p \approx 0.1$, and the above condition is fulfilled.
\begin{figure*}[ht]
  \centering
  \includegraphics[width=0.45\linewidth]{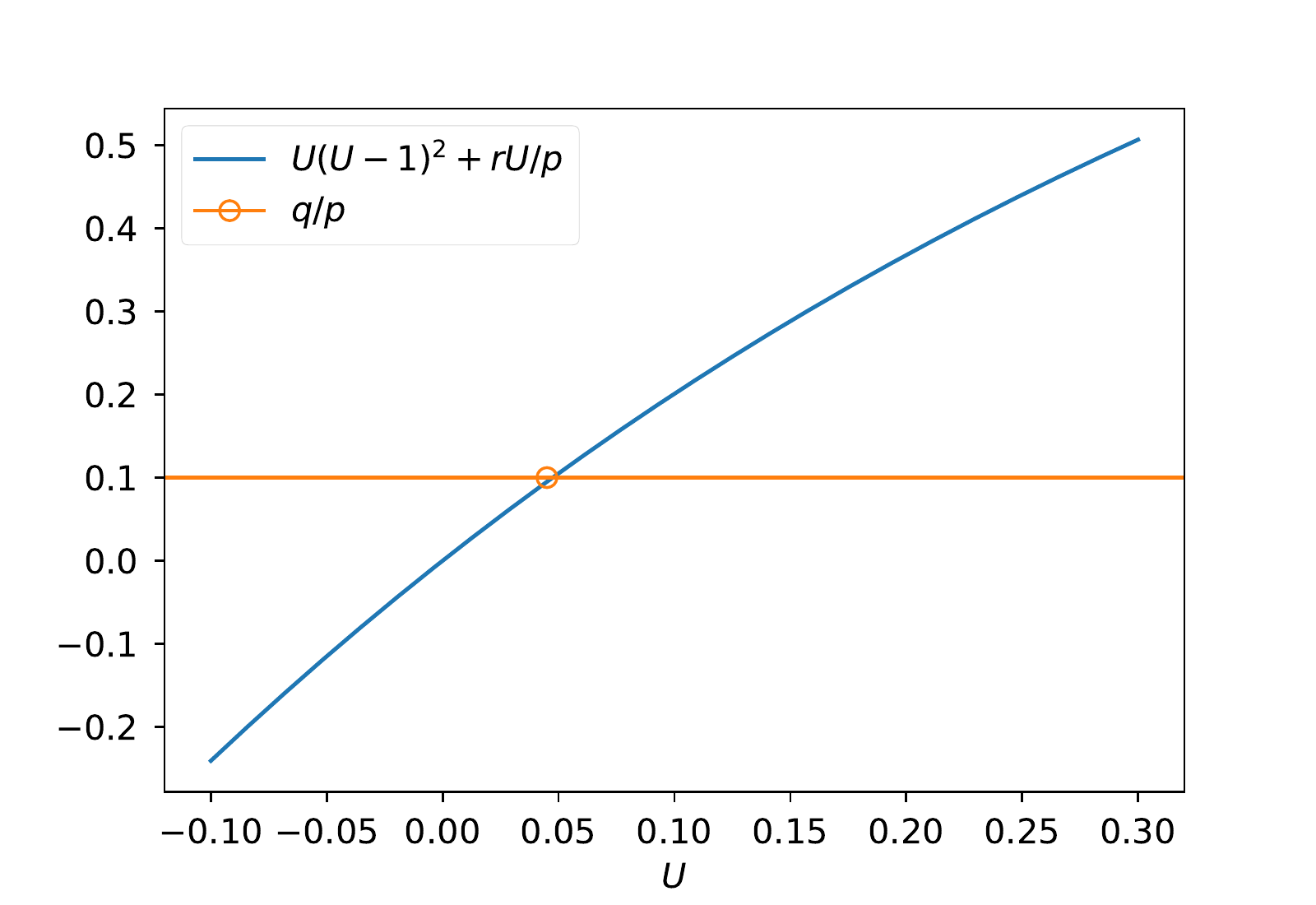}
  \includegraphics[width=0.45\linewidth]{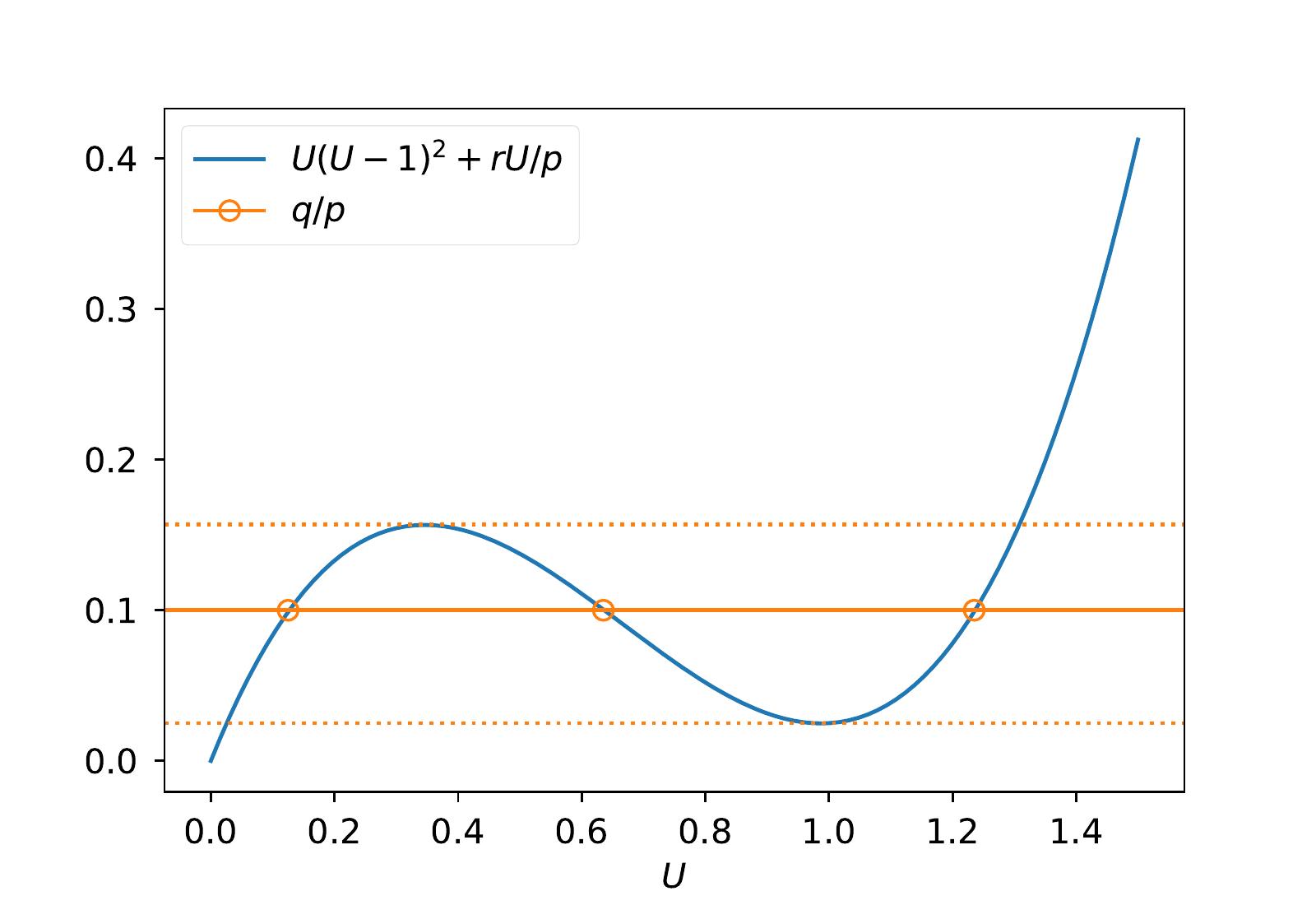}
  \caption{\label{fig:bistab-qualitative-constant} Different terms in the steady-state balance relation~\eqref{eq:simplemomentumbalance}: friction and vertical advection (solid blue) and constant eddy forcing (yellow). Left: $r/p \approx 1$. Right: $r/p = 0.025$. The circles indicate equilibrium states, i.e.\ solutions of the balance equation~\eqref{eq:simplemomentumbalance}.}
\end{figure*}
An illustration is provided in Fig.~\ref{fig:bistab-qualitative-constant}: we plot separately $U{(U-1)}^2+rU/p$ (the sum of vertical advection by the Hadley cell and friction) and the constant forcing $q/p$ for two values of the ratio $r/p$.
When this ratio is small enough ($0 \leq r/p \leq 1/3$), the positive feedback of the Hadley cell leads to the existence of three solutions to Eq.~\eqref{eq:simplemomentumbalance} for some range of forcing amplitude $q/p$ (indicated by the two dashed lines in Fig.~\ref{fig:bistab-qualitative-constant}, right), two stable ones ($U\approx 0.1$ and $U\approx 1.2$ on the figure) and an unstable one ($U \approx 0.6$ on the figure).
As the forcing amplitude sweeps the range of positive values, two (saddle-node) bifurcations are encountered: we start from an equilibrium with weak equatorial wind ($U\approx 0$) for low values of the forcing, which loses stability when the forcing amplitude increases past some value (the dashed line at $q/p \approx 0.16$ in Fig.~\ref{fig:bistab-qualitative-constant}, right).
The system then jumps abruptly to the equilibrium state with strong westerly wind ($U \approx 1.3$) and remains on this branch if the forcing is further increased.
Now, this superrotating equilibrium in turn loses stability when the forcing decreases below some value (the dashed line at $q/p \approx 0.02$ in Fig.~\ref{fig:bistab-qualitative-constant}, right).
We have just described a \emph{hysteresis} phenomenon.
When the ratio $r/p$ becomes too large, the negative feedback of friction overcomes the positive feedback of the Hadley cell, and there is only one solution to Eq.~\eqref{eq:simplemomentumbalance} ($U \approx 0.05$ on the figure) for the whole range of forcing amplitude $q/p$ (Fig.~\ref{fig:bistab-qualitative-constant}, left).

In fact, the eddy forcing $F$ should not be a constant.
In the next section, we show that it may be modelled as a resonant function of $U$, and we discuss in Secs.~\ref{sec:shell}\ref{sec:bistabilitybalancequalitative} and~\ref{sec:shell}\ref{sec:bistabilitybalancequalitative} the consequences for the balance relation~\eqref{eq:simplemomentumbalance}.

\subsection{The wave-jet resonance: Matsuno-Gill computation of the eddy momentum flux convergence}\label{sec:matsuno}

The goal here is to compute the eddy momentum flux convergence induced by a non-zonal tropical heating.
We rely on a classical approach in tropical dynamics: we assume that the zonal mean zonal wind evolves slowly compared to the eddies, and we compute the linear eddy response to the heating term with a constant background wind.
This kind of approach has been used extensively since the pioneering work of~\citet{Matsuno1966} and~\citet{Gill1980}: a major advantage is that for the shallow-water equations on an equatorial beta plane, the linear response can be computed explicitly.
Typically, the stationary response of the atmosphere to a localized heating consists in the superposition of an equatorially trapped Kelvin wave east of the source and a Rossby wave west of the source.
The relative phases of the two standing waves depend on the parameters of the problem.
It has been noticed by several authors~\citep[e.g.][]{Showman2010, Arnold2012} that, in a wide range of parameter values, the Matsuno-Gill response to non-zonal heating could converge westerly momentum at the equator; \citet{Arnold2012} further argued that the response exhibited a resonant structure.
Here, after briefly recalling their result, we compute the associated eddy momentum flux convergence and discuss its resonant structure.

To make the problem analytically tractable, we rewrite Eqs.~\eqref{eq:sweusphere}--\eqref{eq:swehsphere} linearized around a uniform zonal mean-flow $\bar{u}$, using the beta plane approximation~\citep{VallisBook}:
\begin{align}
  \partial_t u' + \bar{u} \partial_x u' - \beta y v'  &= -g \partial_x h' - \epsilon u', \label{eq:sweu}\\
  \partial_t v' + \bar{u} \partial_x v' + \beta y u'  &= -g \partial_y h' - \epsilon v', \label{eq:swev}\\
  \partial_t h' + \bar{u} \partial_x h' + \bar{h} \partial_x u' + \bar{h} \partial_y v' &= Q, \label{eq:swephi}
\end{align}
where $x$ and $y$ represent the zonal and meridional directions, respectively, and $\beta=2\Omega/a$ is the beta effect at the equator (on Earth, $\beta\approx 2.289 \times 10^{-11}$ m\textsuperscript{-1}.s\textsuperscript{-1}).
We have assumed that the background flow has no meridional component ($\bar{v}=0$) and no meridional shear ($\partial_y \bar{u}=0$, $\partial_y \bar{h}=0$).
We are also neglecting the momentum exchange with the underlying layer.
In the rest of this section, we use as time and length units $T=1/\sqrt{\beta c_g}$ and $L=\sqrt{c_g/\beta}$, with $c_g=\sqrt{g\bar{h}}$ the velocity of pure gravity waves.

In the absence of mean-flow ($U=0$), \citet{Matsuno1966} found the normal modes of the linear system~\eqref{eq:sweu}--\eqref{eq:swephi} without forcing and dissipation ($Q=\epsilon=0$), and computed
the stationary solution to the forced-dissipative problem by projecting onto those normal modes.
We refer to~\citet[chap. 8]{VallisBook} or~\citet[chap. 11]{GillBook} for details of the methods, including the dispersion relation and spatial structure of the modes.
The uniform mean-flow $U$ Doppler-shifts the response without modifying the structure of the modes.
For a stationary tropical heating of the form $Q=Q_0\cos(kx)e^{-y^2/4}$, the stationary response consists of two contributions: a Kelvin wave and a Rossby wave with $n=1$, both with zonal wave number $k$~\citep{Arnold2012}.
The expression for the Kelvin mode $(u_K', v_K', h_K')$ reads:
\begin{align}
  u_K' = h_K' &= \frac{-Q_0 \gamma_K}{2\epsilon(1+\gamma_K^2)}\lbrack \gamma_K \cos (kx)+\sin(kx)\rbrack e^{-y^2/4},\\
  v_K' &=0
  \intertext{while the Rossby mode $(u_R', v_R', h_R')$ is given by:}
  u_R' &= \frac{-Q_0 \gamma_R}{6\epsilon(1+\gamma_R^2)}\lbrack \gamma_R \cos (kx)+\sin(kx)\rbrack (y^2-3) e^{-y^2/4},\\
  v_R' &= \Big\lbrace \frac{-4 Q_0 \gamma_R}{3\epsilon(1+\gamma_R^2)}\lbrack (Uk+ \gamma_R \epsilon) \cos (kx) + (\epsilon-Uk\gamma_R)\sin(kx)\rbrack +Q_0\cos(kx)\Big\rbrace y e^{-y^2/4},\\
  h_R' &= \frac{-Q_0 \gamma_R}{6\epsilon(1+\gamma_R^2)}\lbrack \gamma_R \cos (kx)+\sin(kx)\rbrack (y^2+1) e^{-y^2/4},
\end{align}
with $\gamma_X=\epsilon/k(\bar{u}+c_X)$ a non-dimensional parameter defined for the two indices $X=K$ and $X=R$, and $c_X$ the phase velocity of the free waves: $c_R=-1/(3+k^2)$ and $c_K=1$ in non-dimensional units.
The total response is given by $u'=u_R'+u_K', v'=v_R', h'=h_R'+h_K'$.

\begin{figure*}[ht]
  \centering
  \includegraphics[width=0.45\linewidth]{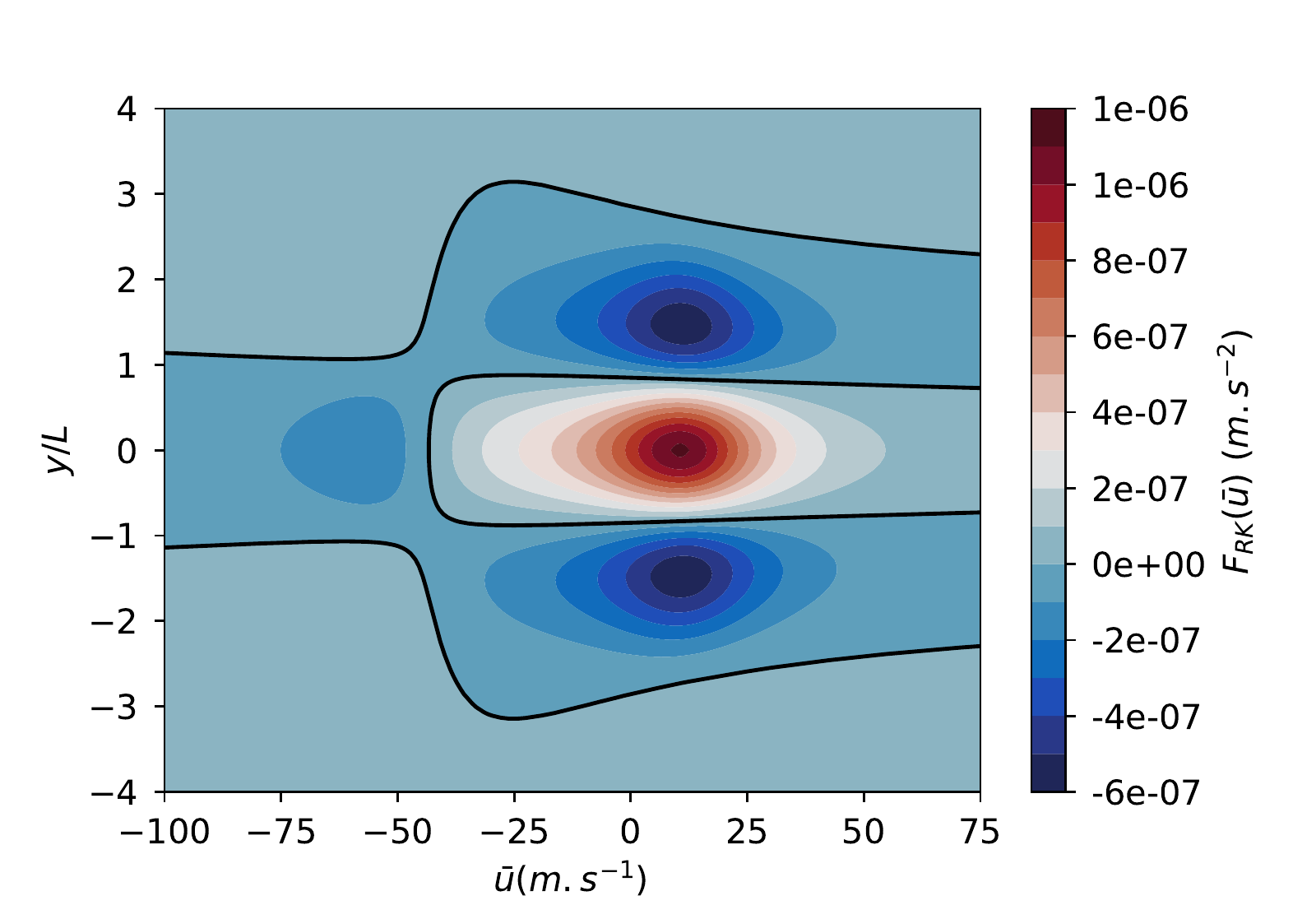}
  \includegraphics[width=0.45\linewidth]{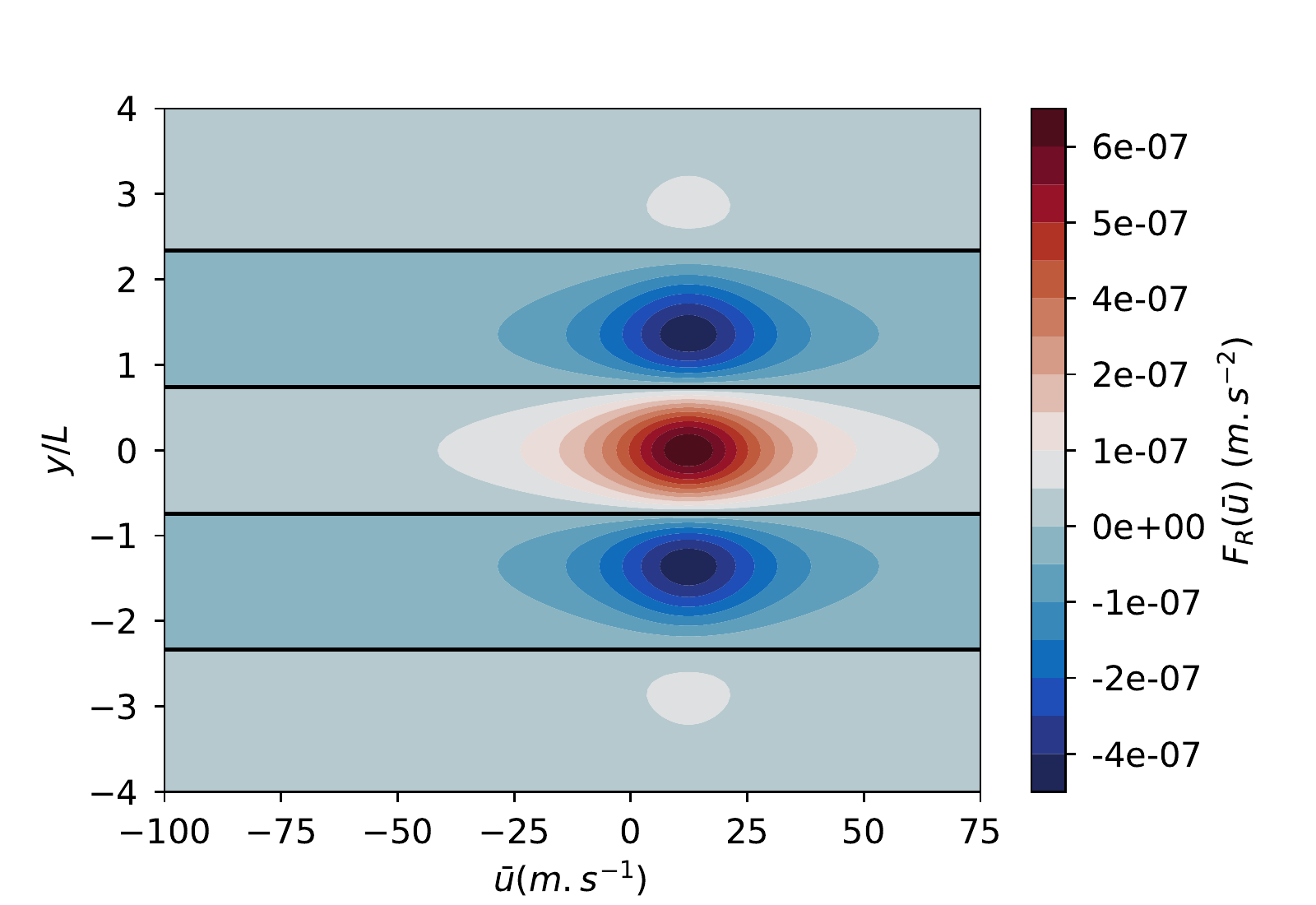}
  \caption{\label{fig:momentumflux-spatial} Contour levels for the eddy momentum flux convergence $F(\bar{u}, y)$ (left) and its contribution from the Rossby mode only ($-\partial_y \langle u_R' v_R'\rangle$, right). The thick black line indicates the null contour.}
\end{figure*}

From this point, an explicit formula can be obtained for the corresponding eddy momentum flux convergence:
\begin{align}
  F(\bar{u}, y) &= -\partial_y \langle u'v' \rangle = -\partial_y \langle (u_R'+u_K')v_R' \rangle,\\
  &= \frac{Q_0^2 \epsilon}{36\lbrack \epsilon^2+k^2{(\bar{u}+c_R)}^2\rbrack} \Big\lbrace \lbrack {(y^2-3)}^2-6 \rbrack + 3\frac{\epsilon^2+k^2{(\bar{u}+c_R)}^2+4k^2c_R(c_K-c_R)}{\epsilon^2+k^2{(\bar{u}+c_K)}^2}(y^2-1)\Big\rbrace e^{-y^2/2},
\end{align}
where the first term in the braces corresponds to the contribution from the Rossby mode only ($-\partial_y \langle u_R' v_R'\rangle$; this corresponds to the case $\gamma_K=0$, or equivalently, $c_K \to +\infty$), and the second term is due to the interaction between the Kelvin and Rossby components of the response ($-\partial_y \langle u_K' v_R'\rangle$).
The spatial structure of the eddy momentum flux convergence $F(\bar{u}, y)$ as a function of the background mean-flow velocity $\bar{u}$, and its contribution from the Rossby mode only ($-\partial_y \langle u_R' v_R'\rangle$), are shown in Fig.~\ref{fig:momentumflux-spatial} in dimensional units.
It is obtained using parameter values $\bar{h}=250$ m, $\epsilon=1$ day\textsuperscript{-1} and $k a=1$.
Going back to the dimensional expression for the phase velocity of the Rossby and Kelvin waves:
\begin{equation}
  c_R = -\frac{\beta}{k^2+(2n+1)\beta/c_g}, \quad c_K=c_g,
\end{equation}
the numerical values for the phase velocities are $c_R \approx -16$ m.s\textsuperscript{-1} (recall that $n=1$) and $c_K\approx 49$ m.s\textsuperscript{-1}.
With these parameters, the Rossby deformation radius is $L \approx 1500$ km.
As expected, the eddy momentum flux convergence is symmetric with respect ot the equator.
For all the values of the background mean-flow $\bar{u}$, the Rossby component (Fig.\ref{fig:momentumflux-spatial}, right) is positive in the equatorial region (within one deformation radius of the equator, roughly speaking, i.e.\ about 13\textdegree), inducing eastward acceleration of the jet, then negative (between one and two deformation radii) and positive again in the extratropics.
A similar spatial structure is found in the full eddy momentum convergence flux (Fig.\ref{fig:momentumflux-spatial}, left), except when the background mean-flow coresponds to strong easterly wind.
In that case, the contour lines are distorted, up to a point where the eddy momentum flux convergence becomes negative in the equatorial region.
Both the full eddy momentum flux convergence and its Rossby component exhibit local maxima and minima, corresponding to resonance and antiresonance structures.
Let us further describe these mechanisms by focusing on the equatorial area.

Let us denote $F_{RK}(\bar{u})$ the full eddy momentum flux convergence at the equator ($y=0$) and $F_R(\bar{u})$ the contribution from the Rossby mode:
\begin{align}
  F_{RK}(\bar{u})  &= F(\bar{u}, 0) = \frac{Q_0^2 \epsilon k^2 (c_K-c_R) (2\bar{u}+c_K -3c_R)}{12\lbrack \epsilon^2+k^2{(\bar{u}+c_R)}^2\rbrack \lbrack \epsilon^2+k^2{(\bar{u}+c_K)}^2 \rbrack}, \label{eq:frk} \\
  F_R(\bar{u})     &= \frac{Q_0^2 \epsilon}{12\lbrack \epsilon^2+k^2{(\bar{u}+c_R)}^2\rbrack}. \label{eq:fr}
\end{align}
It is easily seen from~\eqref{eq:frk} that the eddy momentum flux convergence at the equator $F_{RK}(\bar{u})$ is positive as long as $\bar{u}>(3c_R-c_K)/2$.
$F_R(\bar{u})$, on the other hand, is always positive.
Besides, $F_R(\bar{u})$ has the shape of a Lorentz curve.
\begin{figure}[ht]
  \centering
  \includegraphics[width=\linewidth]{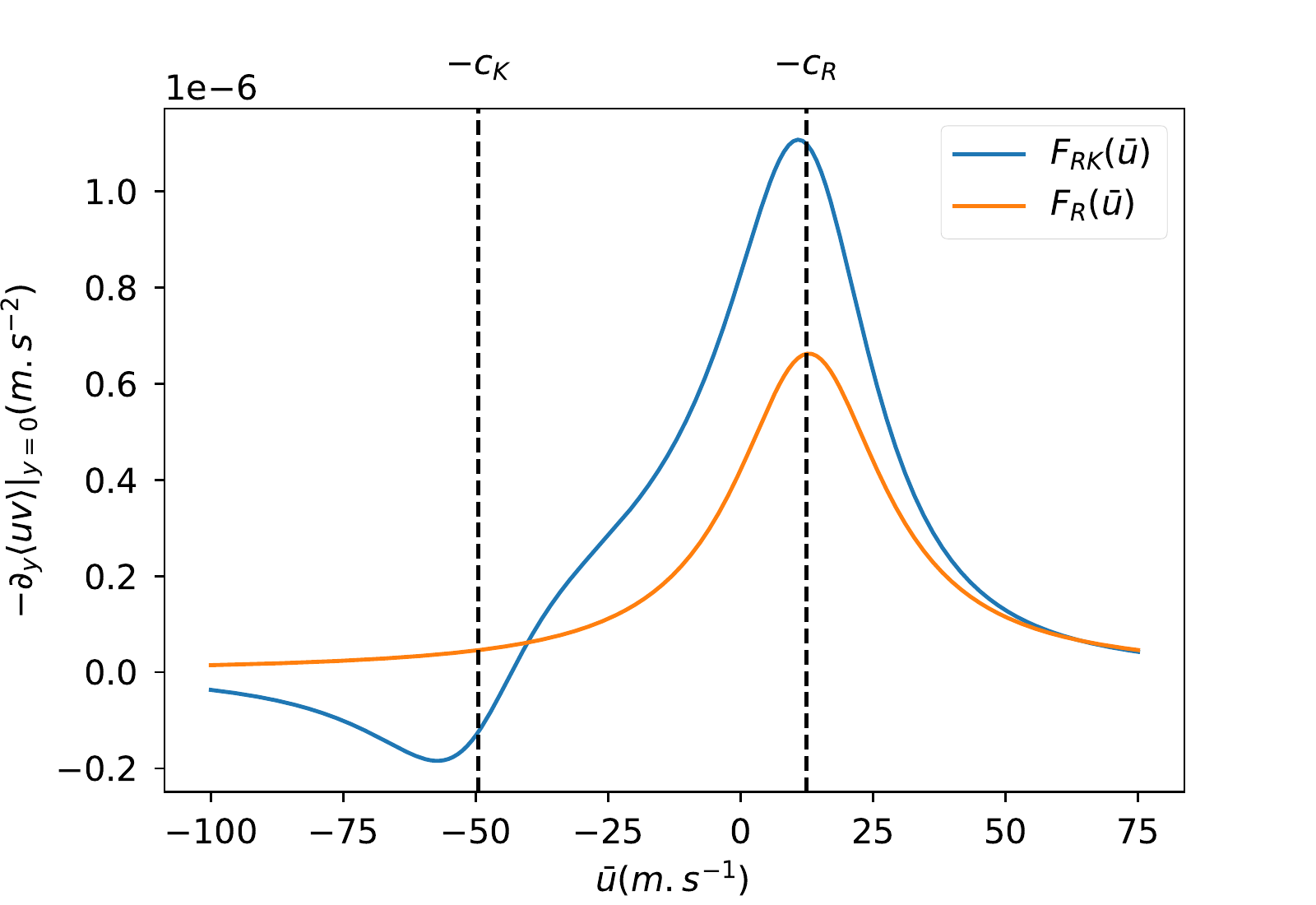}
  \caption{\label{fig:momentumflux} Momentum flux convergence at the equator from the stationary response to the tropical heating, for the Rossby mode (yellow) and the total response (blue). The opposite of the phase velocity of free Rossby and Kelvin waves are indicated with vertical dashed lines.}
\end{figure}
The curves $F_{RK}(\bar{u})$ and $F_R(\bar{u})$ are shown in Fig.~\ref{fig:momentumflux}, using the same parameter values as above.
Both cases exhibit a resonance for background velocities $\bar{u} \approx -c_R$.
When the Kelvin mode is taken into account, there is a secondary peak with opposite sign for $\bar{u} \approx -c_K$.
For the existence of multiple steady-states, a critical point is the sign of the feedback associated to the eddy momentum flux convergence, i.e.\ the sign of the derivative with respect to $\bar{u}$, $\frac{dF_{RK}(\bar{u})}{d\bar{u}}$ or $\frac{dF_R(\bar{u})}{d\bar{u}}$.
From Fig.~\ref{fig:momentumflux}, it is clear that the feedback is positive below the resonance ($\frac{dF_{RK}(\bar{u})}{d\bar{u}} >0$ for $-c_K < \bar{u} < -c_R$) and negative above it ($\frac{dF_{RK}(\bar{u})}{d\bar{u}} <0$ for $\bar{u} > -c_R$).
Ultimately, the existence of multiple steady-states for the mean-flow $\bar{u}$ depends on the other acceleration terms: qualitatively, bistability with a superrotating steady-state hinges on the positive feedback described above overcoming the negative feedbacks due to other effects, such as linear friction for instance (see Sec.~\ref{sec:shell}\ref{sec:bistabilitybalancequalitative}).

Of course, it seems natural that the linear response framework should break down when the amplitude of the forcing becomes too large.
Then, the dynamical feedback of the eddies on the mean flow cannot be neglected anymore.
The linear and nonlinear responses have been compared for instance analytically using perturbative expansion~\citep{Gill1986}, or numerically using idealized models~\citep{NobreThesis} and full GCM simulation~\citep{Lutsko2018}.
In practice however, it has been found in many studies that the linear response computation provides a useful starting point for interpreting results from observations or full nonlinear GCMs~\citep{Moura1981, Gill1983, Neelin1988, Jin1995, Kraucunas2005, Norton2006, Sobel2012, Arnold2012}.
Here, it should be kept in mind that the spatial structure of the response may differ significantly from the linear response in the superrotating state~\citep{Lutsko2018}.
However, most of our reasoning does not depend on the details of the spatial structure, but rather on the resonant behavior which has been reported to hold in a full nonlinear GCM~\citep{Arnold2012} for heating rates and spatial structure similar to those considered here.
Hence, we shall consider that the eddy momentum flux convergence computed in this section is a reasonable working hypothesis, and we shall now study how it may lead to bistability.

\subsection{Qualitative behavior of the wave-jet resonance}\label{sec:bistabilitybalancequalitative}

The functional form of the Rossby wave forcing $F_R(\bar{u})$ (it is a Lorentzian function) makes it simpler than the full resonant eddy forcing $F_{RK}(\bar{u})$, and it also reduces the number of free parameters.
In this section, we exploit this to obtain a qualitative understanding of the steady-states of the momentum budget~\eqref{eq:simplemomentumbalance}.

Injecting~\eqref{eq:fr}, in dimensional units, into the normalized parameters~\eqref{eq:shellparams}, we obtain the corresponding forcing term for the zonal momentum balance model:
\begin{equation}
  q(U) = \frac{\tilde{Q}}{1+\Lambda{(U+c_R/u_{0\text{eq}})}^2},
\end{equation}
with $\tilde{Q}=\beta Q_0^2\tau^2/(12 r u_{0\text{eq}})$ and $\Lambda={(k u_{0\text{eq}}/\epsilon)}^2$, where $k$ is the zonal wave number of the forcing and $\epsilon$ the friction coefficient.
In addition to the parameter $r/p$ discussed in Sec.~\ref{sec:shell}\ref{sec:bistabilityhadley-balance}, which governs the competition between the feedbacks of the two damping mechanisms, vertical advection by the Hadley cell and friction, there are two parameters characterizing the eddy forcing.
First, the position of the resonance is governed by a purely dynamical quantity $-c_R/u_{0eq}$, the phase velocity of free Rossby waves, non-dimensionalized by the velocity associated to the radiative forcing.
Second, the width of the resonance is governed by the parameter $\Lambda$, which depends upon the wave number of the non-zonal forcing, but also the radiative forcing and friction.

\begin{figure*}[ht]
  \centering
  \includegraphics[width=0.45\linewidth]{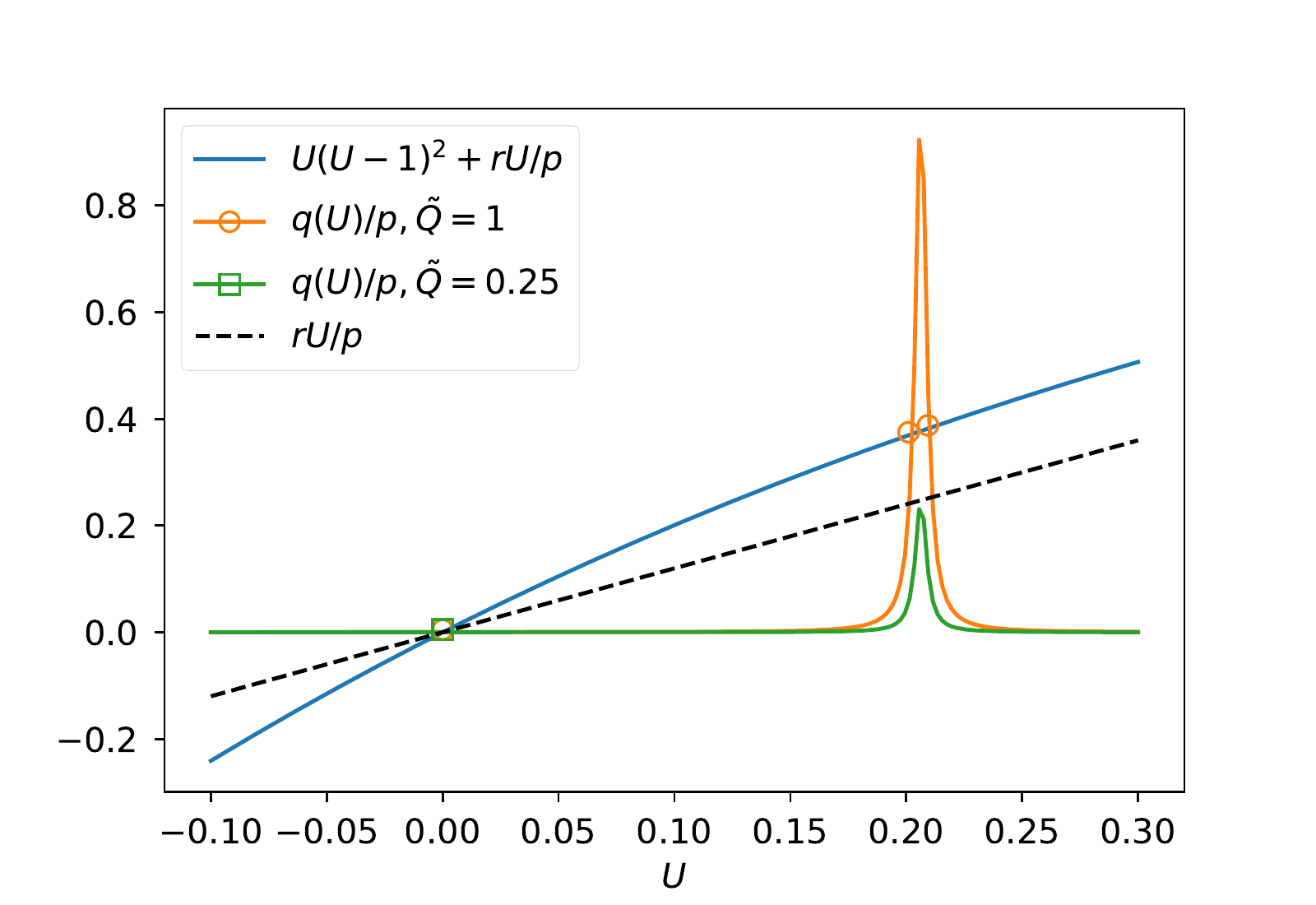}
  \includegraphics[width=0.45\linewidth]{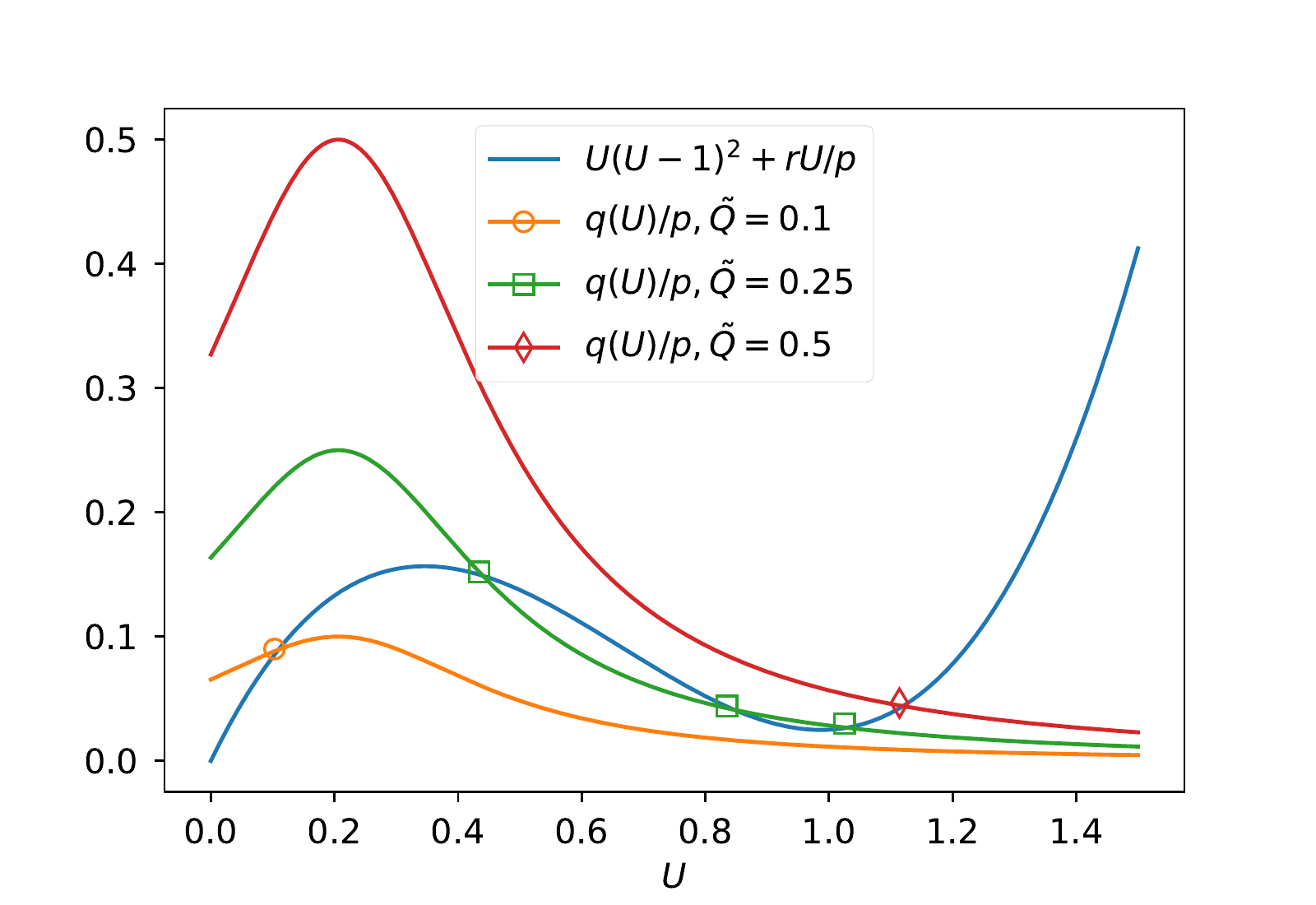}
  \includegraphics[width=0.45\linewidth]{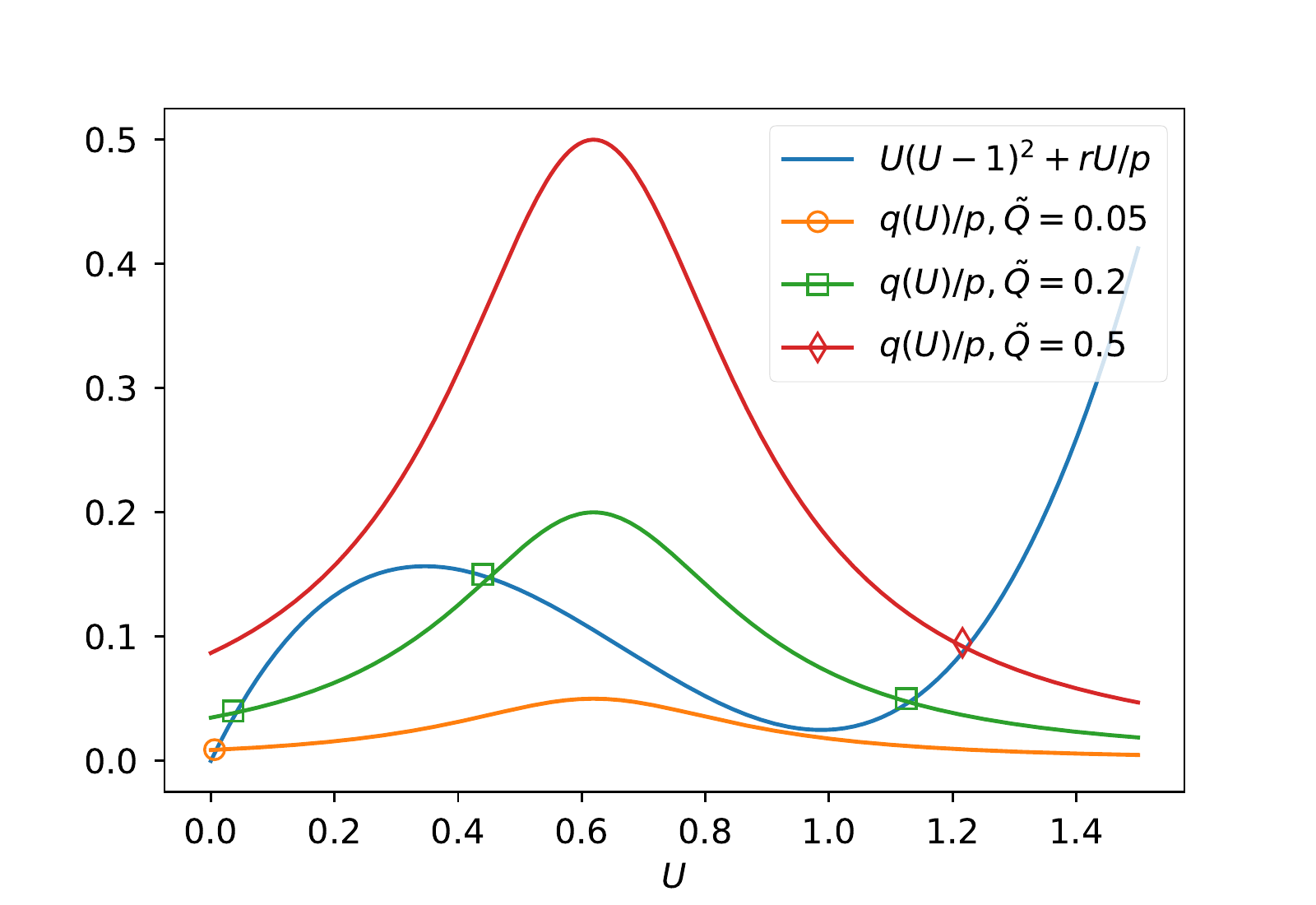}
  \includegraphics[width=0.45\linewidth]{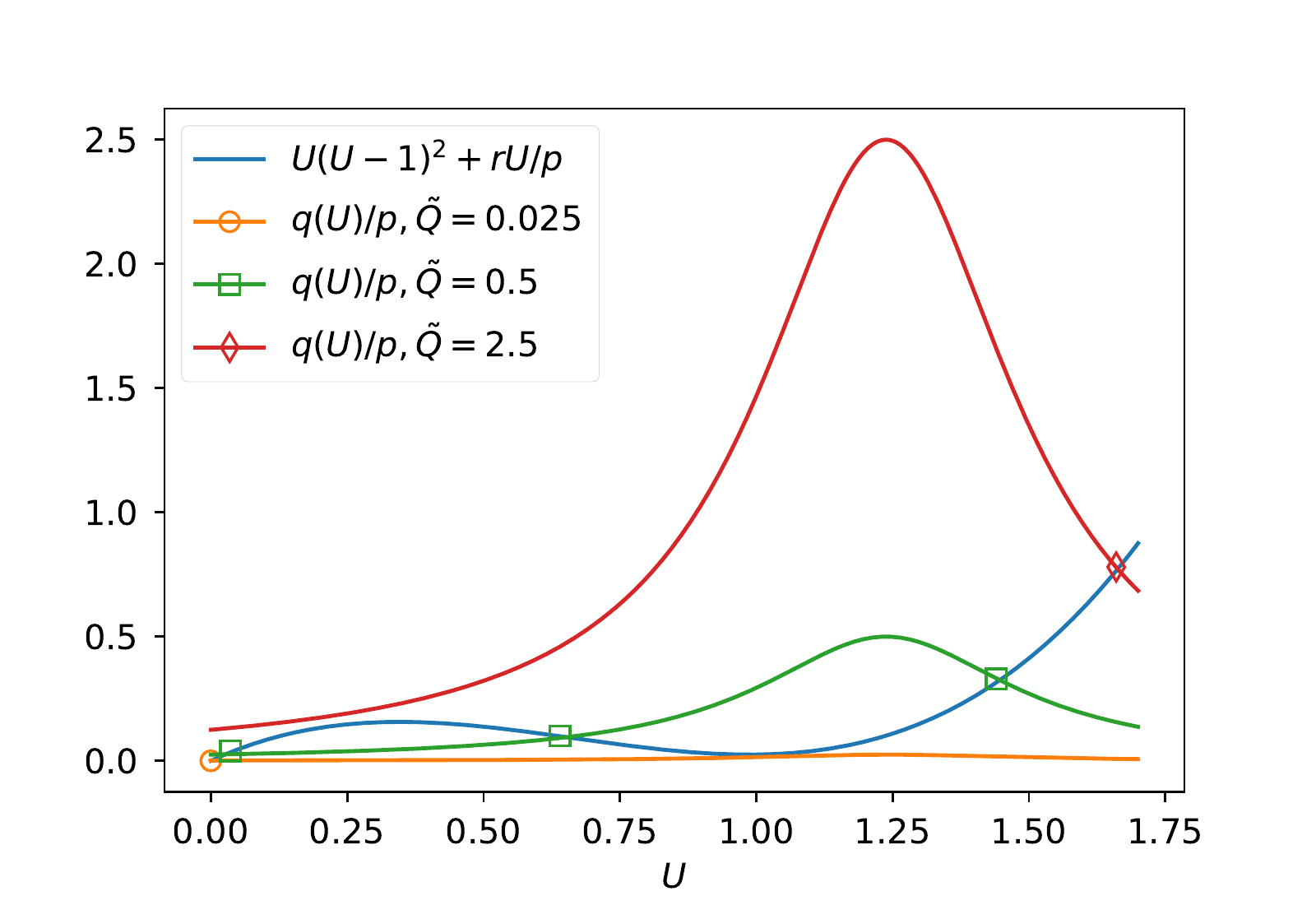}
  \caption{\label{fig:bistab-qualitative} Different terms in the steady-state balance relation~\eqref{eq:simplemomentumbalance}: friction and vertical advection (solid blue) and resonant eddy forcing (colors indicate different forcing amplitudes). Top left: $\Lambda \approx 10^5$, $r/p\approx 1$, $c_R/u_{0eq}=0.2$, the dashed black curve indicates friction alone. Top right: $\Lambda \approx 10$, $r/p=0.025$, $c_R/u_{0eq}=0.2$. Bottom left: $\Lambda \approx 10$, $r/p=0.025$, $c_R/u_{0eq}=0.6$. Bottom right: $\Lambda \approx 10$, $r/p=0.025$, $c_R/u_{0eq}=1.2$. Symbols indicate equilibrium states, i.e.\ solutions of the balance equation~\eqref{eq:simplemomentumbalance}.}
\end{figure*}
Let us discriminate the possibilities for multiple solutions to Eq.~\eqref{eq:simplemomentumbalance} based on the parameter $r/p$:
\begin{itemize}
\item When $r/p> 1/3$, the negative feedback of friction overcomes the positive feedback of the Hadley cell.
  The sum of the two is a monotonously increasing function of $U$.
  For bistability to appear, we need the wave-jet resonance to be sufficiently peaked for the positive feedback due to the eddy forcing to overcome the negative feedback of friction close to the resonance peak.
  This requires that the region with a significant positive feedback (i.e.\ the bump of the Lorentzian) is entirely contained in the $U>0$ range, which can be expressed as $\Lambda \gg {(u_{0eq}/c_R)}^2$.
  This can be checked explicitly by setting $R=0$ in the simplified zonal momentum balance, which yields the equation $q=rU$: this is a cubic equation which can be solved exactly.
  In this case, a bistability range appears as soon as $\Lambda > 3$.
  Then, provided the forcing amplitude is large enough, there are three solutions to the balance equation: an unstable one and two stable ones.
  We refer to this case as \emph{resonance-driven} bistability: it is illustrated in Fig.~\ref{fig:bistab-qualitative} (top left).
  One of the stable states correspond to $U\approx 0$ --- on the left flank of the resonance peak ---, and the other one is a superrotating state, with $u_0 \approx -c_R$ (for an infinitely narrow resonance) --- on the right flank of the resonance peak.
  A first saddle-node bifurcation occurs when $\tilde{Q}$ increases and the resonance peak intersects the friction curve, corresponding to the appearance of the superrotating state.
  A second saddle-node bifurcation occurs when the forcing becomes significantly non-zero for $U$ close to zero, corresponding to the loss of stability of the conventional circulation.
  However, this second bifurcation is expected to occur for very large forcing amplitudes: in other words, the range of forcing amplitude for which bistability occurs should be very wide in this scenario.
  Note that the stable superrotating state is very close to the unstable state.

\item When $r/p < 1/3$, there is a region where the positive feedback of the Hadley cell acts.
  Multiple steady-states may also exist as the eddy forcing amplitude varies, depending on the other parameters.
  Indeed, for an infinitely wide resonance ($\Lambda \ll 1$), we should recover the case studied in Sec.~\ref{sec:shell}\ref{sec:bistabilityhadley-balance}, governed entirely by the Hadley cell feedback.
  For a resonant eddy forcing with finite width, let us distinguish three cases, based on the position of the resonance, for a fixed value of $r/p$.
  \begin{itemize}
  \item  When the resonance peak occurs at a velocity smaller than the local maximum of $U{(U-1)}^2+rU/p$ (Fig.~\ref{fig:bistab-qualitative}, top right), the same kind of scenario as in the previous paragraph unfolds, except that in the regime where three equilibria exist, they are all on the right flank of the resonance peak, i.e.\ in the region where the eddy forcing feedback is negative.
    This means that this time, bistability relies on the feedback from the Hadley cell.
    It should also be noted that this type of bistability is more difficult to obtain with the eddy forcing than with a constant forcing, as was considered in Sec.~\ref{sec:shell}\ref{sec:bistabilityhadley-balance}.
    Fig.~\ref{fig:bistab-qualitative} shows that this scenario occurs already for $\Lambda \approx 10$.

  \item When the resonance peak occurs at a velocity between the local maximum and the local minimum of $U{(U-1)}^2+rU/p$ (Fig.~\ref{fig:bistab-qualitative}, bottom left), bistability is again possible.
    This time, the two stable states are always on different flanks of the resonance peaks, while the unstable one moves from the right flank to the left flank as the forcing amplitude increases (until it annihilates with the low wind stable state at the saddle-node bifurcation).
    In other words, the appearance of the superrotating state occurs because the positive feedback of the Hadley cell sets in, like in the previous case, but, on the other hand, the loss of stability of the conventional state is due to the positive wave-jet feedback prevailing over the negative feedback of the Hadley cell.
  \item When the resonance peak occurs at a velocity larger than the local minimum of $U{(U-1)}^2+rU/p$, (Fig.~\ref{fig:bistab-qualitative}, bottom right).
    The first saddle-node bifurcation, corresponding to the appearance of the superrotating case, occurs on the left flank of the resonance peak.
    As the forcing amplitude keeps increasing, the superrotating state moves to the right flank of the lorentzian.
    In this case both feedbacks contribute with the same sign.
  \end{itemize}

\end{itemize}

As shown in Fig.~\ref{fig:momentumflux}, the eddy momentum flux convergence associated to the full response (i.e.\ including the projection on the Kelvin mode) is amplified compared to the Rossby mode response, but the overall structure remains qualitatively similar (if we except the negative tail for strong easterly background winds).
Hence, the analysis of the limiting cases carried out above is not expected to change qualitatively when including the Kelvin contribution.

\subsection{Quantitative discussion}\label{sec:bistabilitybalancequantitative}

In Sec.~\ref{sec:shell}\ref{sec:bistabilitybalancequalitative}, we have considered independently the role of the three non-dimensional parameters ($r/p$, $\Lambda$ and $c_R/u_{0eq}$) characterizing the balance between zonal acceleration due to resonant eddy forcing, vertical advection by the Hadley cell and friction.
We have given simple criteria for bistability due to the Hadley cell feedback ($r/p \leq 1/3$) and the wave-jet resonance ($\Lambda \gg {(u_{0eq}/c_R)}^2$, i.e. $k^2c_R^2/\epsilon^2 \gg 1$).
Let us now illustrate this balance for some typical parameter values.

\begin{figure*}[ht]
  \centering
  \includegraphics[width=\linewidth]{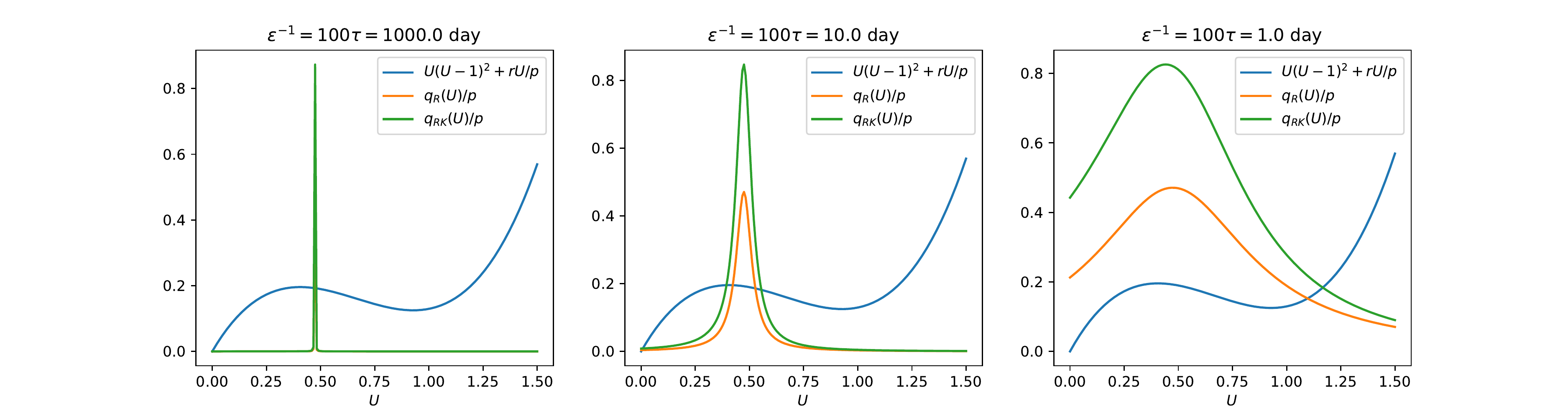}
  \caption{\label{fig:bistab-shell} The different terms of the balance relation~\eqref{eq:simplemomentumbalance} as functions of $U$ (sum of momentum exchange with the underlying layer and Rayleigh friction in blue, eddy forcing due to Rossby --- resp. Rossby-Kelvin --- modes in yellow --- resp.\ green), varying the friction time $\epsilon^{-1}$ (decreasing from left to right) such that the ratio with the Newtonian cooling time scale $\tau$ is kept constant (the other parameters are taken from Table~\ref{tab:shellheld}). With these parameters, the value of $\Lambda$ in decreasing order is roughly 10\textsuperscript{7}, 10\textsuperscript{3} and 10.}
\end{figure*}

We first consider the parameter values from~\citet{Shell2004}, summarized in Table~\ref{tab:shellheld}, supplemented with forcing parameters $k a=1$ and $c_R=-16$ m.s\textsuperscript{-1}.
Such values fall under the scenario where there is bistability, governed by the resonant eddy feedback because the resonance is very strongly peaked ($c_R^2/(\epsilon^2 L^2) \approx 5\times 10^5$), as can be seen by plotting the terms of the balance relation~\eqref{eq:simplemomentumbalance} like in the above (Fig.~\ref{fig:bistab-shell}, left).
However, the value used for friction is lower than typical values for the atmosphere of the Earth, by several orders of magnitude (about 0.001 day\textsuperscript{-1}, instead of 0.1--1 day\textsuperscript{-1}, e.g.~\citet{Held1994}).
As explained by~\citet{Shell2004}, this is essentially a consequence of the simplistic vertical structure of the model.
In reality, dissipative processes modelled by linear friction have a more complex physical nature (eddy viscosity, wave breaking, etc).
Before considering a refined description of the vertical structure of the atmosphere (see Sec.~\ref{sec:heldhou}), let us comment briefly on the effect of variations of the friction coefficient in this simple model.
It is clear from the definition of the non-dimensional parameters that the regime of the Hadley cell remains the same if we increase $\epsilon$ while decreasing simultaneously the radiative cooling time $\tau$ (one could equivalently decrease the layer thickness at the equator, $h_{0eq}$).
Fig.~\ref{fig:bistab-shell} (middle and right) shows that varying friction across a range of typical values, while keeping its ratio with the radiative cooling timescale constant, exhibits a transition from the scenario governed by the wave-jet feedback to the scenario governed by the Hadley cell.
\begin{table*}
  \centering
\begin{tabular}{ccccccc}
              & $\beta$ (m\textsuperscript{-1}.s\textsuperscript{-1}) & $c_g$ (m.s\textsuperscript{-1}) & $L$ (km) & $c_R$ (m.s\textsuperscript{-1}) & $\epsilon$ (day\textsuperscript{-1}) & $c_R^2/(\epsilon^2 L^2)$ \\
  \hline
  Earth       & $2.289 \times 10^{-11}$ & 50 & 1500  & 16  & 0.1 & 90 \\
  Jupiter     & $5 \times 10^{-12}$     & 680 & 11500 & 230 & 0.002 & $7.2\times 10^{5}$ \\
              &                         &     &       &     & 0.05  & 1150\\
  Hot Jupiter & $7.8 \times 10^{-13}$   & 2000 & 50000 & 590 & 0.1 & 100 \\
              &                         &      &       &     & 1   & 18
\end{tabular}
\caption{Parameters for different planetary atmospheres: The Earth, Jupiter (two values of $\epsilon$ from~\citet{Warneford2017} and~\citet{Schneider2009}), and Hot Jupiter exoplanets~\citep{Showman2011} such as HD189733b.}\label{tab:parameters}
\end{table*}
In addition, we list in Table~\ref{tab:parameters} estimates of parameter values for different planetary atmospheres, which indicate that the bistability regime governed by the wave-jet feedback seems relevant in most cases of interest.
However, these estimates hinge crucially on the friction coefficient $\epsilon$, which is difficult to estimate for the reasons mentioned above.
Hence, investigations with a more realistic model are necessary.
Before doing so (Sec.~\ref{sec:heldhou}), let us comment on differences between the two bistability regimes in the simple model.

\begin{figure*}[ht]
  \centering
  \includegraphics[width=0.45\linewidth]{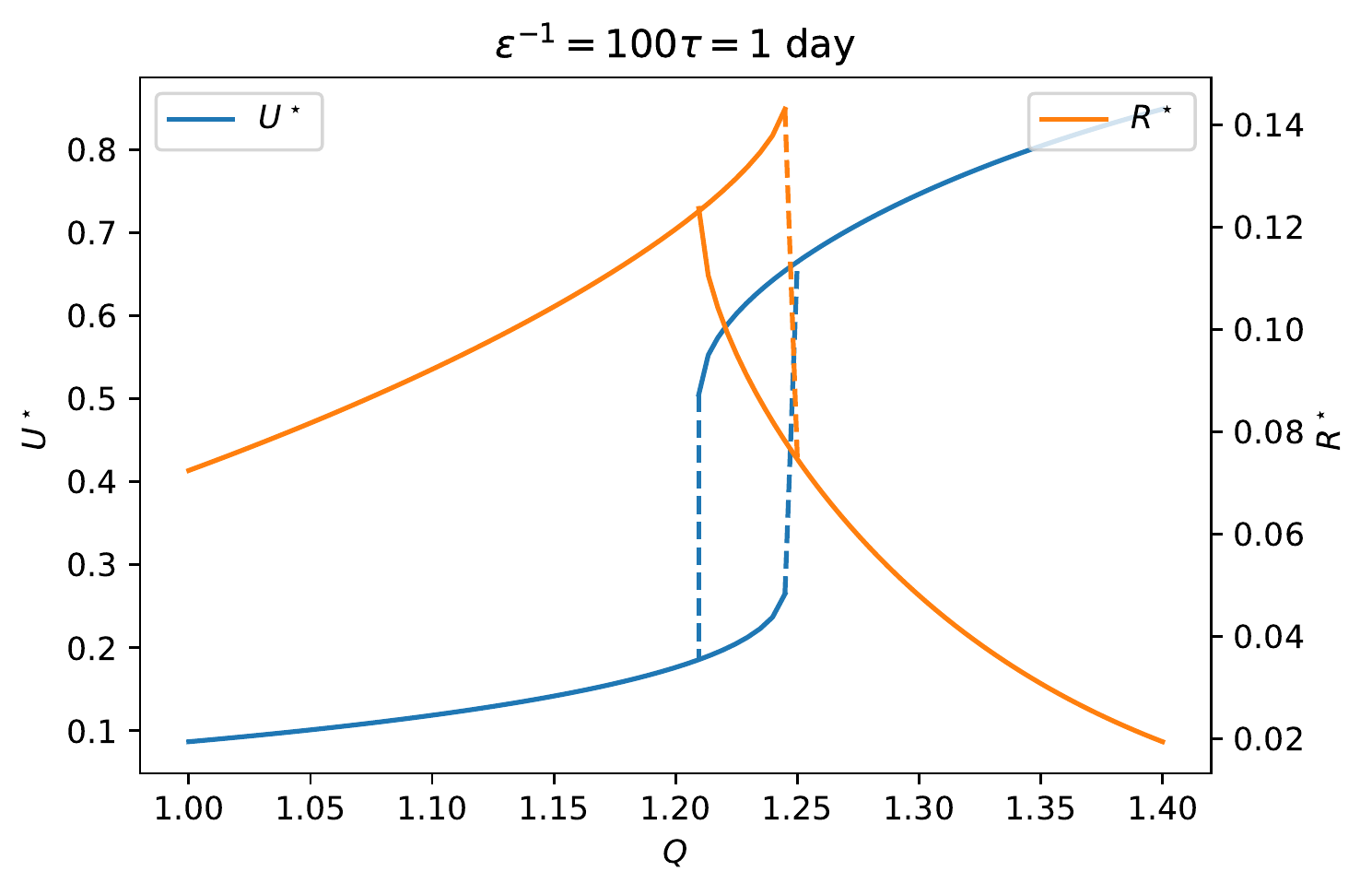}
  \includegraphics[width=0.45\linewidth]{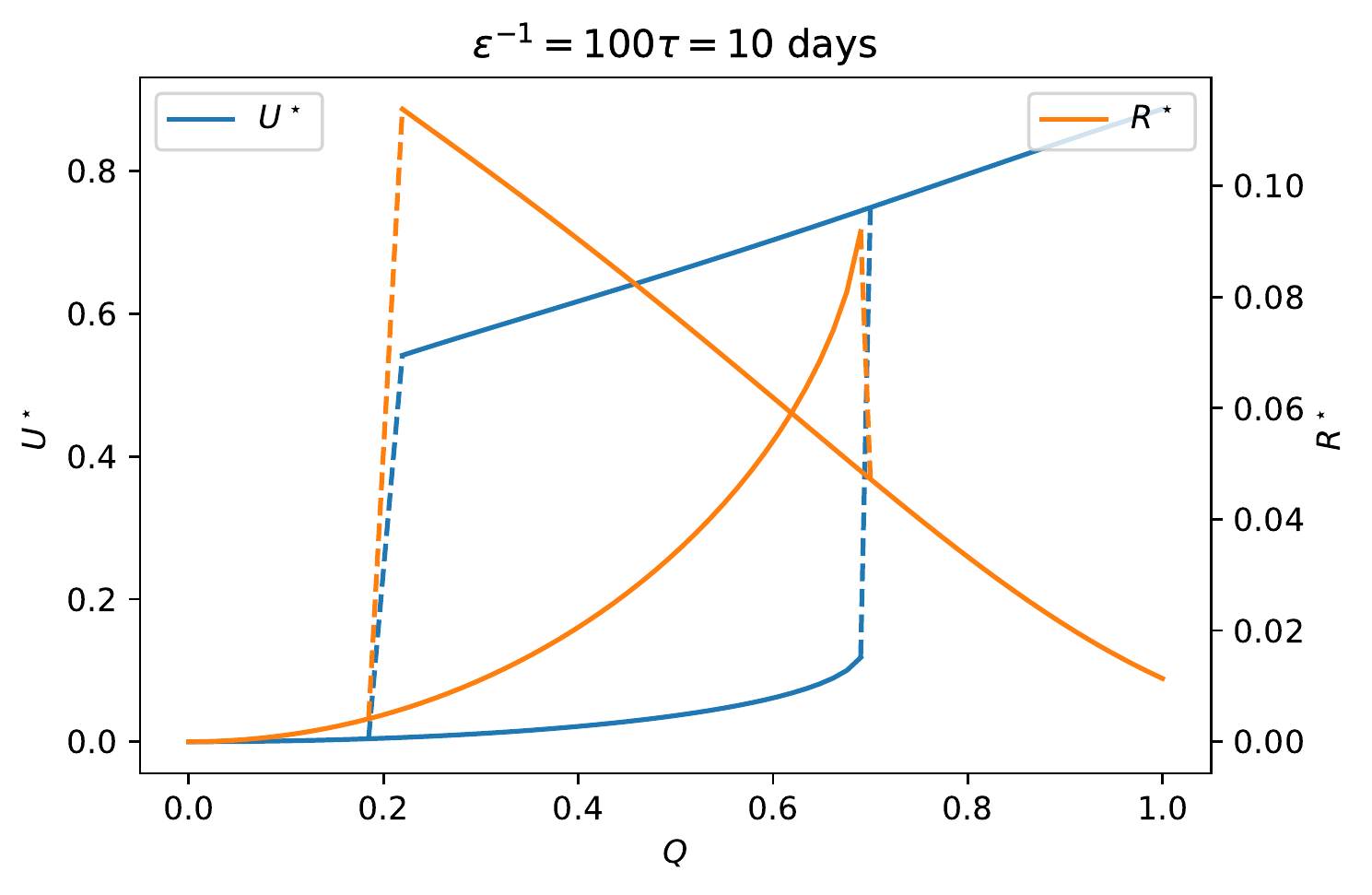}
  \caption{\label{fig:hysteresis} Hysteresis curves for the equilibrium equatorial zonal wind $U^*$ and vertical momentum advection $R^*$ for the two cases: bistability governed by the Hadley cell feedback (left) and by the resonant eddy forcing (right).}
\end{figure*}
The hysteresis curves obtained by tracking the solution of the balance equation~\eqref{eq:simplemomentumbalance} as we ramp up and down the forcing amplitude, both for velocity $U$ and vertical advection of zonal momentum $R$, are shown in Fig.~\ref{fig:hysteresis}.
We show the same figure for two cases: one where bistability is governed by the Hadley cell feedback ($\epsilon=1$ day\textsuperscript{-1}, Fig.~\ref{fig:hysteresis}, left), and one where bistability is governed by the resonant eddy forcing feedback ($\epsilon= 0.1$ day\textsuperscript{-1}, Fig.~\ref{fig:hysteresis}, right).
As anticipated in the qualitative study, while both cases exhibit bistability, the bistability range is much wider in the case dominated by the resonant eddy forcing.
The behavior of the Hadley cell is also quite different in the two cases: it collapses in the superrotating state governed by the Hadley cell feedback, but this is not necessarily the case in the superrotating case induced by the resonant eddy forcing.

\section{Bistability in the axisymmetric primitive equations}\label{sec:heldhou}

\subsection{Numerical setup}

We now investigate the interplay between the resonant eddy forcing and the Hadley cell feedbacks in a more realistic context.
Instead of the zonally-averaged shallow-water equations (Eq.~\eqref{eq:swezonalwind} for the zonal wind), we consider the axisymmetric primitive equations:
\begin{align}
  \frac{\partial u}{\partial t} + \frac{v}{a}\frac{\partial u}{\partial \phi} + \omega \frac{\partial u}{\partial p} - \frac{uv \tan \phi}{a}  &= 2 \Omega v \sin \phi + F_u + \nabla \cdot \tau_u, \label{eq:primitiveaxiu}\\
  \frac{\partial v}{\partial t} + \frac{v}{a}\frac{\partial v}{\partial \phi} + \omega \frac{\partial v}{\partial p} + \frac{u^2 \tan \phi}{a}  &= -2 \Omega u \sin \phi - \frac{1}{a} \frac{\partial \Phi}{\partial \phi} + F_v + \nabla \cdot \tau_v, \label{eq:primitiveaxiv} \\
  \frac{\partial \theta}{\partial t} + \frac{v}{a}\frac{\partial \theta}{\partial \phi} + \omega \frac{\partial \theta}{\partial p} &= - \frac{\theta-\theta_e}{\tau} + \nabla \cdot \tau_\theta, \label{eq:primitiveaxitheta}\\
  \frac{\partial \omega}{\partial p} &= - \frac{1}{a \cos \phi} \frac{\partial }{\partial \phi} (v\cos \phi), \label{eq:primitiveaxiw}\\
  \frac{\partial \Phi}{\partial p} &= - \frac{R T}{p}, \label{eq:primitiveaxiphi}
\end{align}
where the zonally-averaged zonal and meridional wind $u$ and $v$ are now 2D fields (depending on latitude $\phi$ and pressure $p$), $\omega=Dp/Dt$ is the zonally-averaged vertical velocity in pressure coordinates, $\Phi$, $T$ and $\theta$ are the zonally-averaged geopotential, temperature and potential temperature.
This model was considered for instance by~\citet{Held1980} to build a theory of the Hadley cell analogous to the one rapidly presented in Sec.~\ref{sec:shell}\ref{sec:bistabilityhadley-balance}.
Dissipative effects are represented generically by the zonal and meridional components of the zonal-mean stress tensor, $\tau_u$ and $\tau_v$.
$F_u,F_v$ represent the divergence of the Reynolds stress tensor, i.e.\ the eddy forcing.
In our numerical simulations, we prescribe the eddy forcing to account for the wave-jet resonance in a simplified manner.
The only diabatic heating term is a Newtonian relaxation term which drives the temperature field towards a prescribed radiative-convective equilibrium: $\theta_{e}(p,\phi) = \max\left(200(p_0/p)^{R/c_p} , \theta_\star - \Delta_h \sin^2\phi - \Delta_v \ln(p/p_0) \cos^2\phi \right)$.
We use standard values for the coefficients~\citep{Held1994}: $\theta_\star = 315 K$, $\Delta_h = 60 K$, $\Delta_v = 10K$.
The relaxation time $\tau$ is as in~\citet{Held1994}.

The main difference with the shallow-water model considered in Sec.~\ref{sec:shell} is a more accurate description of the vertical structure of the atmosphere, which allows to properly resolve vertical momentum transport by the Hadley cell, through the term $\omega\partial_p u$.

The model is solved numerically using the \emph{Climt} framework~\citep{Caballero2008, Monteiro2018}, which solves the above equations in flux form, using a simple upwind scheme~\citep{Smolarkiewicz1983}.
We use 91 grid points in latitude (i.e.\ a resolution slightly smaller than 2\textdegree{}), and 45 vertical levels (unless specified otherwise).
A turbulent diffusion scheme is used for the stress tensor $\tau$; we shall denote $\nu$ the kinematic viscosity (in m\textsuperscript{2}.s\textsuperscript{-1}) in the vertical direction.
Our runs use the value $\nu=0.5$ m\textsuperscript{2}.s\textsuperscript{-1} by default.
Surface momentum drag is parameterized through a bulk aerodynamic formula~\citep{Caballero2008}, akin to the linear friction considered above.

We carry out two series of numerical experiments, corresponding to two different kinds of prescribed eddy forcing:
\begin{itemize}
\item A resonant eddy forcing $F_u=F_{RK}(u(\phi=0))$ with spatial structure given by the Matsuno-Gill problem (Sec.~\ref{sec:shell}\ref{sec:matsuno}) and with a varying amplitude given by Eq.~\eqref{eq:frk} (see Fig.~\ref{fig:momentumflux-spatial}, left).
  These experiments are designed to reproduce the behavior observed in GCM studies with non-zonal tropical heating, such as~\citet{Suarez1992, Saravanan1993, Kraucunas2005, Arnold2012}.
  Since the model considered here is axisymmetric, we need to parameterize the effect of the eddy forcing, for which we use the analytical computation of the linear response of a shallow-water atmosphere to a non-zonal tropical heating carried out in Sec.~\ref{sec:shell}\ref{sec:matsuno}.
  This allows us to explore parameter space at a much lower computational cost.
\item A constant eddy forcing $F_u$ with the same spatial structure as above and fixed amplitude $F_{RK}(U=0)$.
  These experiments amount to adding a vertical dimension to the setup of~\citet{Shell2004} (the meridional structure is also slightly different).
\end{itemize}
In both cases, the vertical structure is arbitrarily chosen as a Gaussian profile $e^{-(p-p_0)/2\sigma^2}$ centered on the $p_0 = 300$ hPa level, as in~\citet{Caballero2018}.
In these experiments, we always have $F_v=0$.
Note also that we do not parameterize eddy heat transport.

For both types of experiments, we shall be interested in steady-state solutions of the 2D axisymmetric primitive equations~\eqref{eq:primitiveaxiu}--\eqref{eq:primitiveaxiphi}.
Specifically, we want to know whether superrotating solutions exists, and whether multiple solutions may coexist for some values of the forcing parameters.
In particular, we shall vary the resonance width parameter (in the case of the resonant eddy forcing) to illustrate the occurrence of both kinds of bistability identified in Sec.~\ref{sec:shell}.
We shall also discuss the role of viscosity $\nu$ and the vertical resolution.

\subsection{Control run}

Before investigating bistability, let us first show a control run without eddy forcing ($F_u=0$).
\begin{figure}[ht]
  \centering
  \includegraphics[width=\linewidth]{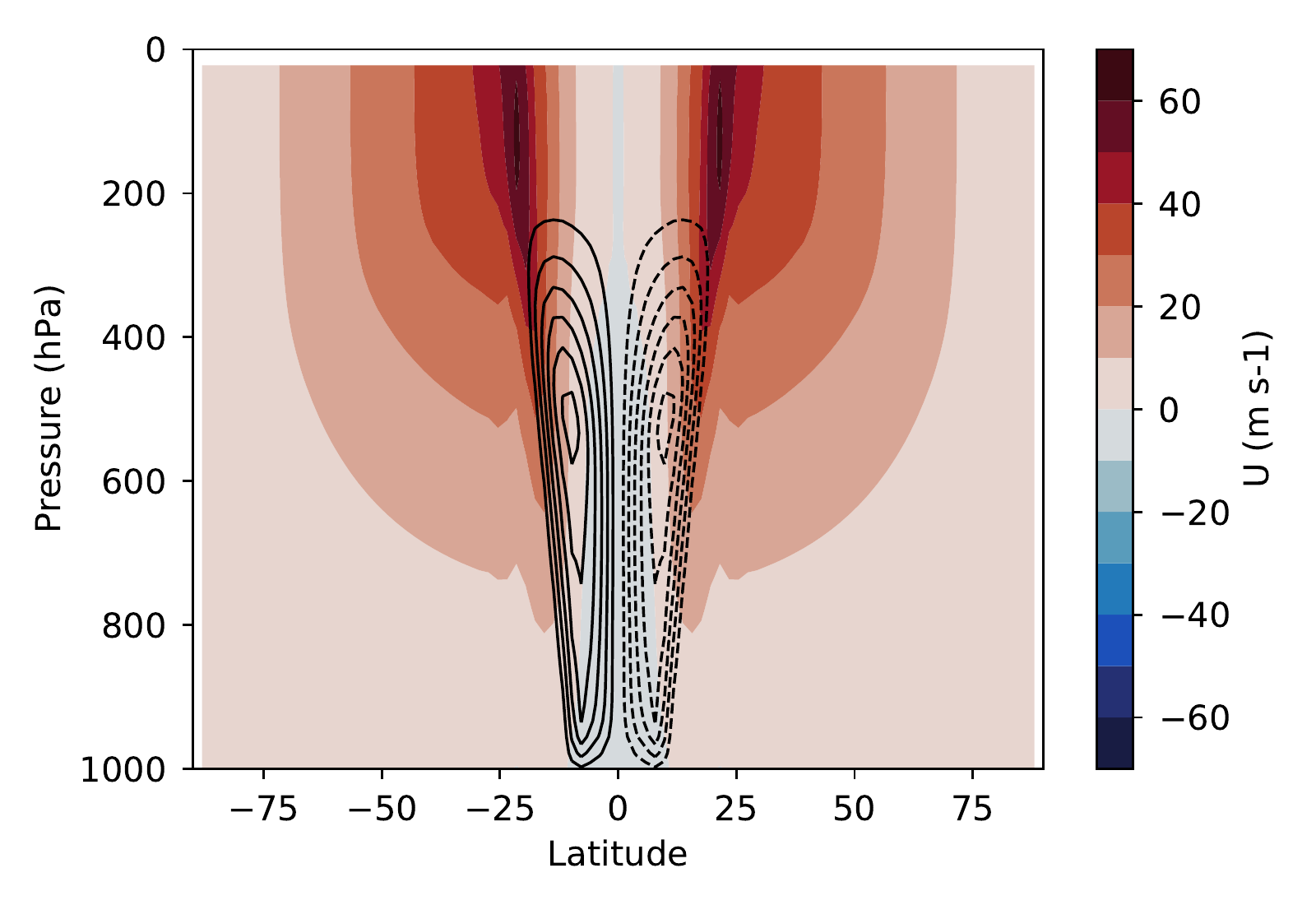}
  \caption{\label{fig:controlu} Zonal wind field (shading) and meridional mass streamfunction (contours; contour interval is 10\textsuperscript{3} m.Pa.s\textsuperscript{-1}, negative contours are dotted) in the control run ($F_u=0$).}
\end{figure}
The equilibrium zonal wind field is shown in Fig.~\ref{fig:controlu}.
Jets (with maximum wind speed $\approx 60$ m.s\textsuperscript{-1}) are obtained in each hemisphere at the poleward edge of the Hadley cell, which extends approximately to 20\textdegree{} in both hemispheres.
Easterly winds prevail in the tropical regions; in particular at the equator, the wind is easterly at all levels.
This control run does not exhibit superrotation.

A more realistic control run could be obtained by prescribing additional eddy momentum (or heat) forcing in the mid-latitudes~\citep{Schneider1984, Singh2016}, as was done in~\citet{Caballero2018}.
For simplicity, we prefer not to do so here.

\subsection{Resonance-driven bistability}\label{sec:resonancebistability}

In a first set of experiments, using the resonant eddy forcing, we illustrate the type of hysteresis identified in Sec.~\ref{sec:shell} where bistability is driven by the resonant response to the forcing.

%
%
\begin{figure}[ht]
  \centering
  \includegraphics[width=\linewidth]{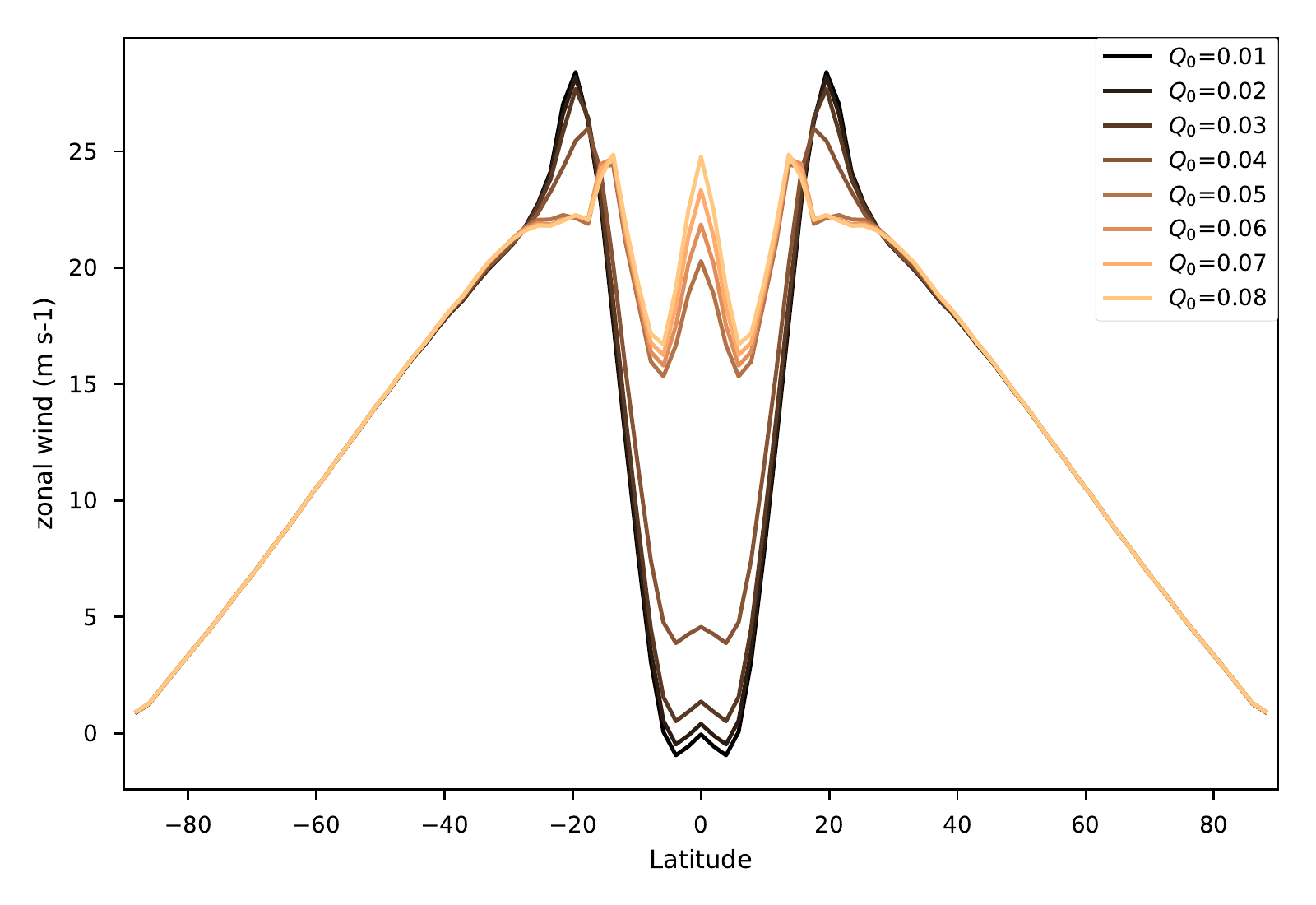}
  \caption{\label{fig:zonalwindpfl_resonance} Vertically averaged zonal wind profile for the resonant eddy forcing with width $\epsilon=0.1$ for different forcing amplitudes.}
\end{figure}
First, we integrate the axisymmetric primitive equations until a statistically stationary state is reached (typically about 1500 days).
We show in Fig.~\ref{fig:zonalwindpfl_resonance} the vertically averaged zonal wind profile at steady-state for a narrow resonance ($\epsilon=0.1$), as the forcing amplitude $Q_0$ is varied.
For low values of the forcing amplitude, the zonal wind profile is essentially fixed by angular momentum conservation in the tropics and radiative equilibrium outside.
Generally speaking, this state has similar characteristics as the control run: full spatial structure of the zonal wind field, mean meridional circulation,\dots
In particular, it exhibits jets close to 20\textdegree{} latitude, as we have seen in the control run.
As the forcing amplitude $Q_0$ increases, these jets move equatorward and weak westerlies appear in the tropics.
When $Q_0$ further increases, there is a relatively sharp transition (between $Q_0 = 0.4$ and $Q_0 = 0.5$) to a different regime where a jet appears on the equator, which quickly becomes as strong as the subtropical jets.
In this regime, the atmosphere is clearly in a state of equatorial superrotation.
The full spatial structure of the wind field in the conventional state is similar to the control run, shown in Fig.~\ref{fig:controlu}.
The circulation in the superrotating state, shown in the right panel of Fig.~\ref{fig:comparison}, will be discussed in more details in Sec.~\ref{sec:heldhou}\ref{sec:comparison}.

%
%
We now carry out hysteresis experiments to investigate the possibility that the conventional and superrotating states coexist in some range of forcing amplitude.
The experiment consists in increasing the forcing amplitude step by step and letting the system relax to its new equilibrium state at each step.
This introduces a discontinuity (in time) in the forcing, but it allows for clearer diagnostics of the response of the system.
Typically, we observe a smooth relaxation to a new equilibrium state, possibly with an initial overshoot.
As expected, relaxation to the new steady-state upon application of the step forcing takes longer close to the bifurcation points.
To ensure that the system has relaxed, we choose a time interval between two steps several times longer than the typical relaxation time observed in previous runs.
We apply this procedure up to a given value of the forcing amplitude (larger than the amplitude threshold for which we observe the abrupt transition to superrotation in the steady-state experiments above), then we reverse the procedure by decreasing the forcing amplitude step by step until we reach the initial forcing amplitude.
Any observable can then be computed as a function of time, or equivalently as a function of the forcing amplitude, with the only difference that in the latter case, it may take one value on the way up and a different one on the way down.

\begin{figure}[ht]
  \centering
  \includegraphics[width=\linewidth]{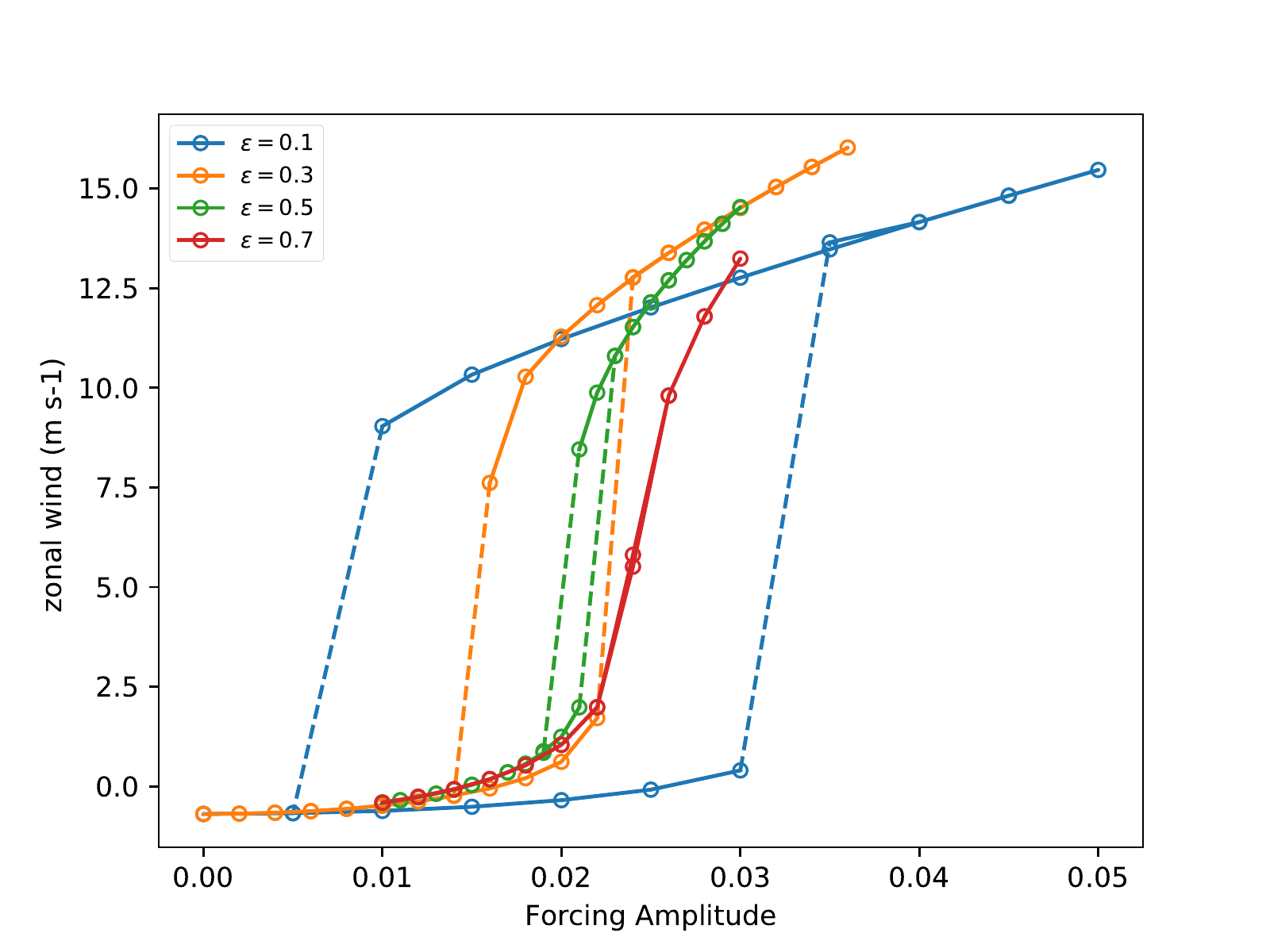}
  \caption{\label{fig:hysteresis_resonance} Hysteresis curves for vertically averaged (between 200 and 700 hPa) zonal wind at the equator (averaged between 5\textdegree{} S and 5\textdegree{} N), for varying resonance width parameter $\epsilon$.}
\end{figure}
The results of the hysteresis experiments are shown in Fig.~\ref{fig:hysteresis_resonance}, for different values of the parameter $\epsilon$.
The observable plotted in the figure is the zonal wind, averaged over a range of latitude around the equator (here between 5\textdegree{} S and 5\textdegree{} N) and over the upper troposphere (here between 200 and 700 hPa).
For small values of $\epsilon$ (narrow resonance, e.g. $\epsilon=0.1$), the averaged zonal wind, initially negative, first increases slowly when the forcing amplitude is increased, then abruptly switches to a positive value (above 10m.s\textsuperscript{-1}), characteristic of a superrotating state.
This corresponds to the behavior observed with the steady-states experiments in the above paragraph, and suggests that the conventional circulation becomes unstable (saddle-node bifurcation).
Once in the superrotating state, the averaged zonal wind again increases slowly with the forcing amplitude until the maximum value of the forcing amplitude is reached.
When the forcing amplitude is decreased, the averaged zonal wind decreases slowly, down to a forcing amplitude below the critical point where the conventional circulation became unstable.
Then, a second bifurcation occurs: the superrotating state becomes unstable and the averaged zonal wind suddenly switches back to its value in the conventional circulation.

The forcing amplitudes corresponding to the bifurcation points depend on $\epsilon$.
More precisely, the bistability range decreases significantly as $\epsilon$ is increased (see the curves for $\epsilon=0.3$ and $\epsilon=0.5$), i.e.\ as the resonance broadens, as anticipated in Sec.~\ref{sec:shell}.
When $\epsilon$ is sufficiently large (e.g\ $\epsilon=0.7$), the bifurcation points disappear entirely: the upper and lower branch of the hysteresis curves collapse onto a single curve, describing the smooth growth of the averaged zonal wind with the forcing amplitude.

\subsection{Hadley cell-driven bistability}\label{sec:hadleybistability}

We now turn to the second series of runs with a constant eddy forcing.
Since the resonance mechanism is manually switched off in this case, the only possibility for bistability to occur is through the Hadley cell feedback.
The goal is to investigate whether the bistability due to this feedback mechanism, obtained analytically in the simple zonal wind balance model of Sec.~\ref{sec:shell} and observed in numerical simulations of the 1-1/2 layer shallow-water equations~\citep{Shell2004}, subsists in our multi-layer configuration.

\begin{figure}[ht]
  \centering
  \includegraphics[width=\linewidth]{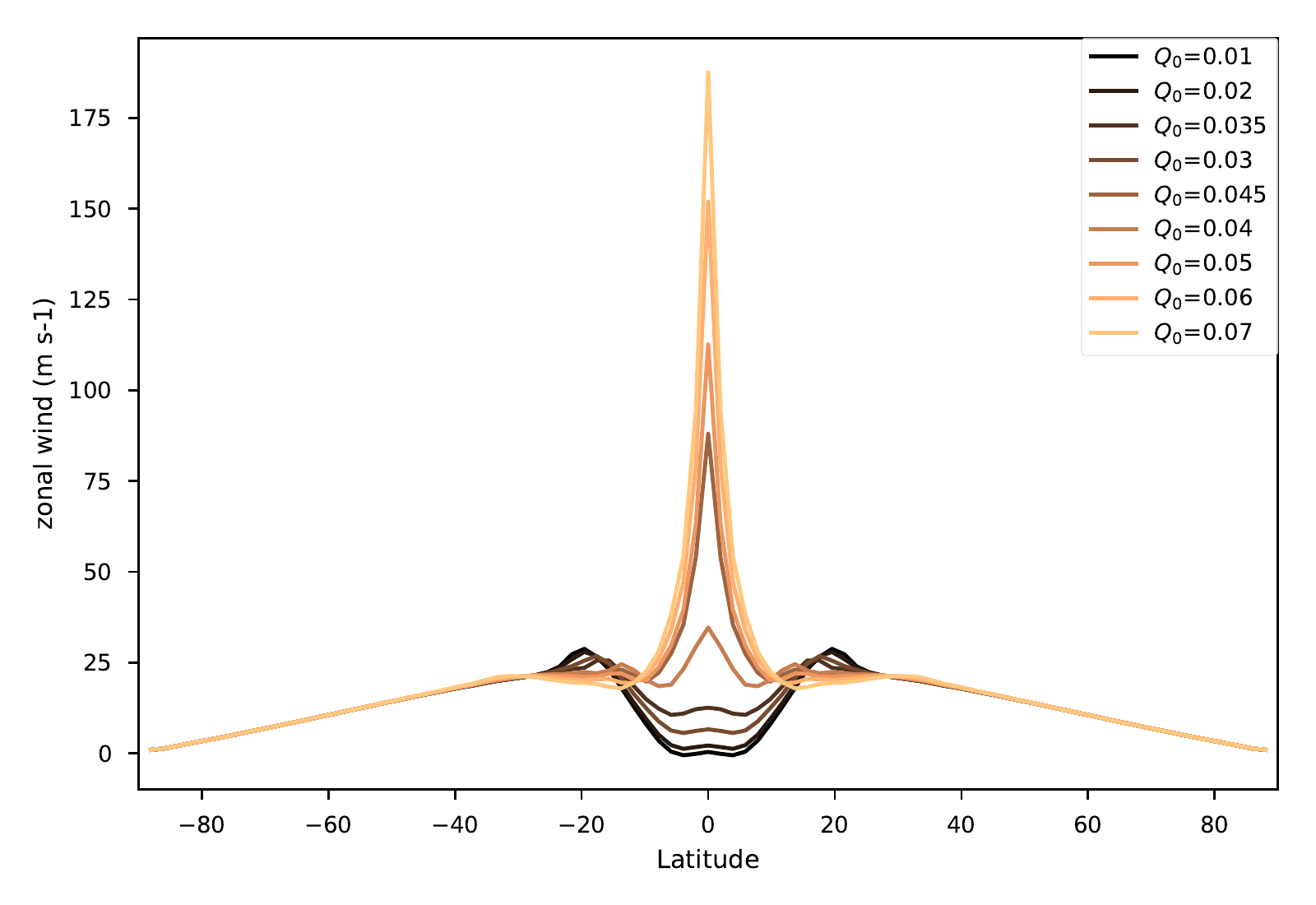}
  \caption{\label{fig:zonalwindpfl_hadley} Vertically averaged zonal wind profile for the constant eddy forcing for different forcing amplitudes.}
\end{figure}
Like in Sec.~\ref{sec:heldhou}\ref{sec:resonancebistability}, we start by studying the steady-states of the axisymmetric equations: the typical relaxation time from an initial state of rest is similar to the resonant eddy forcing (about 1500 days), although the larger values of $Q_0$ require longer integrations (up to about 4500 days).
%
%
Fig.~\ref{fig:zonalwindpfl_hadley} shows the vertically averaged zonal wind obtained in the steady-state for many forcing amplitudes $Q_0$.
For both kinds of forcings, the qualitative behavior is similar to the one described in Sec.~\ref{sec:heldhou}\ref{sec:resonancebistability}.
For low values of the forcing amplitude, the zonal wind profile is similar to the control run, with subtropical jets at the poleward edge of the Hadley cell.
As the forcing amplitude increases, there is a sharp transition to a superrotating circulation.
The value of the threshold amplitude is similar in both forcing cases ($Q_0 \approx 0.04 $), although the magnitude of the equatorial wind in the superrotating state is much larger with the constant forcing.

\begin{figure}[ht]
  \centering
  \includegraphics[width=\linewidth]{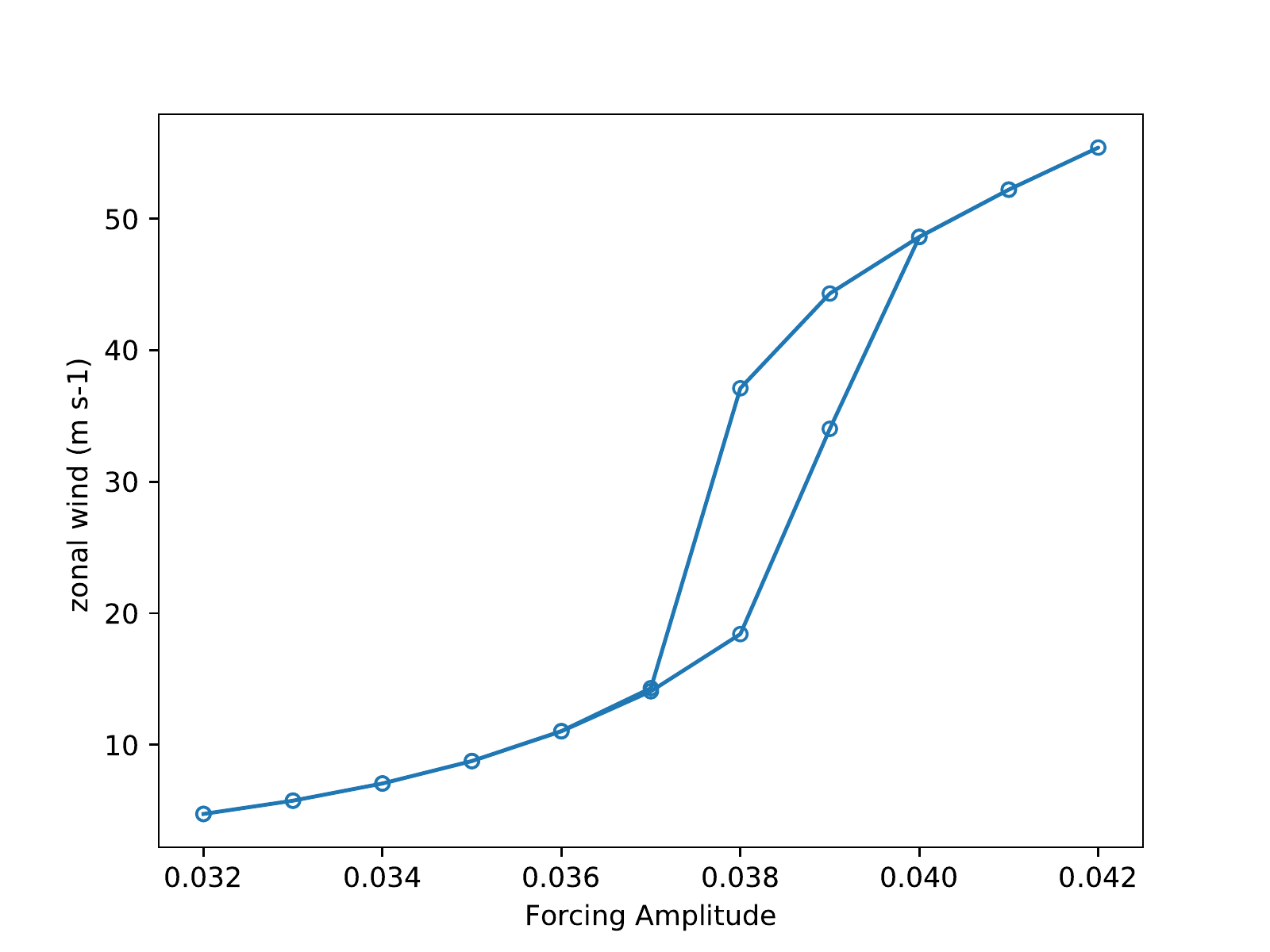}
  \caption{\label{fig:hysteresis_hadley} Hysteresis experiment showing the vertically averaged zonal wind at the equator as a function of forcing amplitude for a constant eddy forcing.}
\end{figure}

%
%
We now carry out a hysteresis experiment following the same protocol as in Sec.~\ref{sec:heldhou}\ref{sec:resonancebistability} (Fig.~\ref{fig:hysteresis_hadley} shows the averaged zonal wind for this hysteresis experiment): we increase step by step the forcing amplitude, allowing the system to relax to its new steady-state at each time, until an abrupt transition to a superrotating state is found.
Then we further increase the forcing amplitude to show that the averaged zonal wind increases smoothly on the superrotating branch, before reverting the loop.
We decrease the forcing step by step, until the superrotating state loses stability, and the averaged zonal wind abruptly goes back to the values found on the way up.
This shows that the constant eddy forcing also exhibits bistability, as was found in the simple analytical model (Sec.~\ref{sec:shell}) and in a single-layer shallow water model~\citep{Shell2004}.
However, it should be noted that the bistability range is much smaller than in the case of resonance-driven bistability studied in Sec.~\ref{sec:heldhou}\ref{sec:resonancebistability}.

\begin{figure}[ht]
  \centering \includegraphics[width=\linewidth]{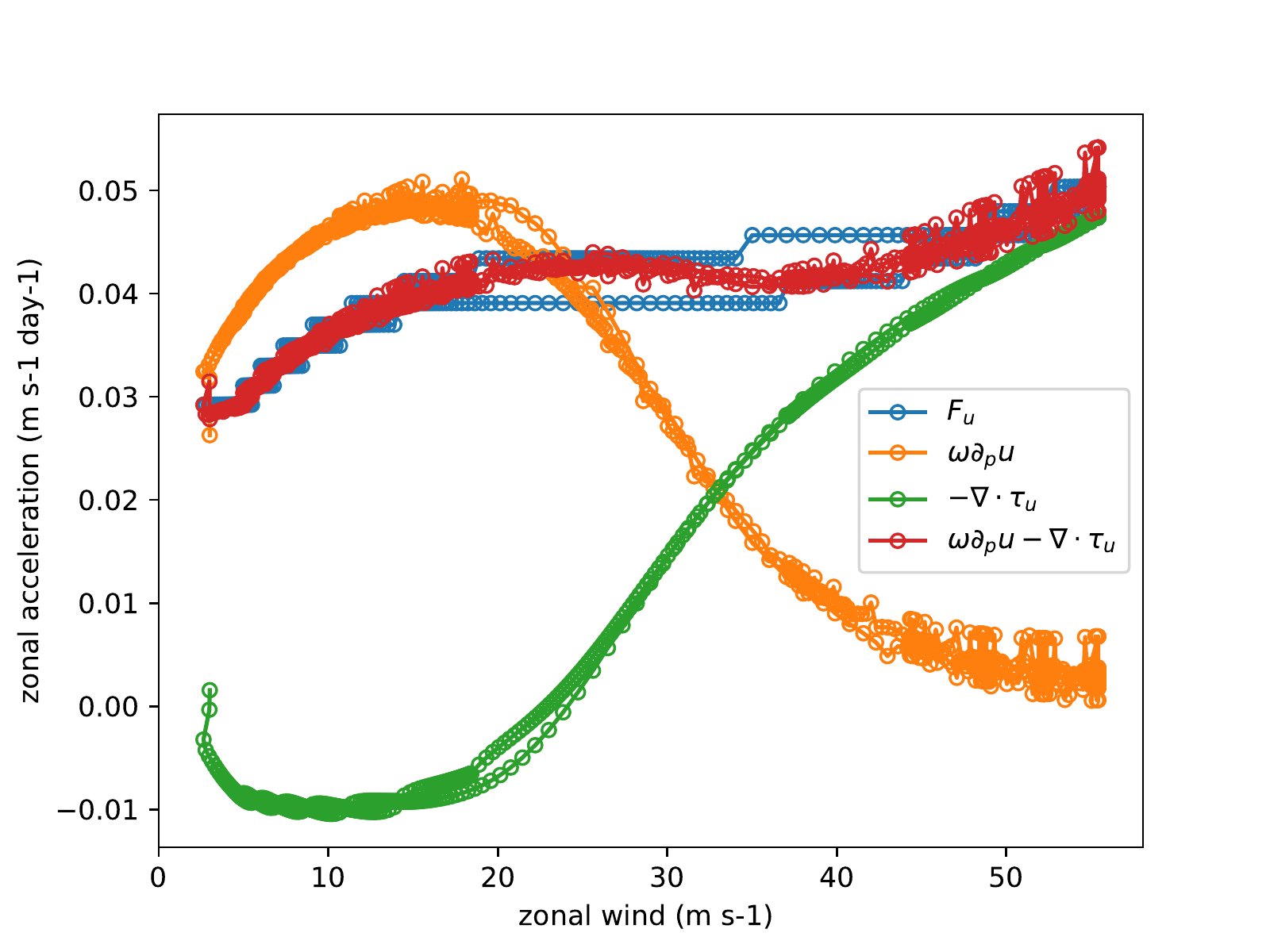}
  \caption{\label{fig:ubudgetglobal_hadley} Zonal acceleration budget, averaged over the tropical upper troposphere, in the hysteresis experiment for the constant forcing case.}
\end{figure}
Like in the zero-dimensional model of zonal momentum balance studied analytically in Sec.~\ref{sec:shell}, we can diagnose the zonal acceleration budget in our axisymmetric simulations.
We show in Fig.~\ref{fig:ubudgetglobal_hadley} the three dominant terms: the eddy forcing (blue curve) as well as the vertical advection of zonal momentum by the Hadley cell, $\omega \partial_p u$ (orange curve) and the turbulent momentum diffusion term $\nabla \cdot \tau_u$ (green curve).
While the former term is prescribed, the latter terms are dynamically adjusted.
These curves are constructed by plotting these terms as functions of the zonal wind, both quantities being averaged over the tropical upper troposphere, in the hysteresis experiment shown in Fig.~\ref{fig:hysteresis_hadley}.
In particular, it contains points which correspond to transient states.
The vertical transport by the Hadley cell exhibits the same kind of cubic behavior as in the analytical model shown in Fig.~\ref{fig:bistab-qualitative-constant}.
Since the dissipative mechanism is not linear friction, the turbulent momentum diffusion curve is not just a straight line, but it is nevertheless an increasing function of the local zonal wind, apart from very low values of the wind.
Both mechanisms act essentially as damping effects (again, except for the lower values of the zonal wind as far as turbulent diffusion is concerned).
We also display the sum of the two effects as a separate curve (red curve): for these parameter values, there exists a range of wind velocities where the positive feedback of the Hadley cell prevails over the negative feedback of the eddy viscosity, and the net damping is not a monotonous function of the zonal wind.
Hence, the qualitative behavior is the same as in Fig.~\ref{fig:bistab-qualitative-constant}: steady-state solutions of the zonal momentum budget should equilibrate this net damping by a prescribed eddy forcing, which is a straight horizontal line in this case.
For a fixed forcing amplitude in a given range, bistability may occur.
In the figure, we show the prescribed eddy forcing in the hysteresis experiment, where the forcing amplitude is time-dependent.
This shows that the hysteresis experiment explores successive steady-states over the two increasing branches of the net damping curve.
The decreasing branch of the net damping curve can only be seen because we have included values from transient states.

\subsection{Numerical convergence with vertical resolution}

Bistability driven by the Hadley cell feedback had previously been observed in an axisymmetric model representing only one vertical mode~\citep{Shell2004}.
In Sec.~\ref{sec:heldhou}\ref{sec:hadleybistability}, we have found that it subsists only marginally in a multilevel setup: there is still bistability but the range of coexistence of the conventional and superrotating states is very narrow.
To better understand this behavior, we have performed steady-state and hysteresis experiments with various vertical resolutions.

\begin{figure*}[ht]
  \centering
  \includegraphics[width=0.45\linewidth]{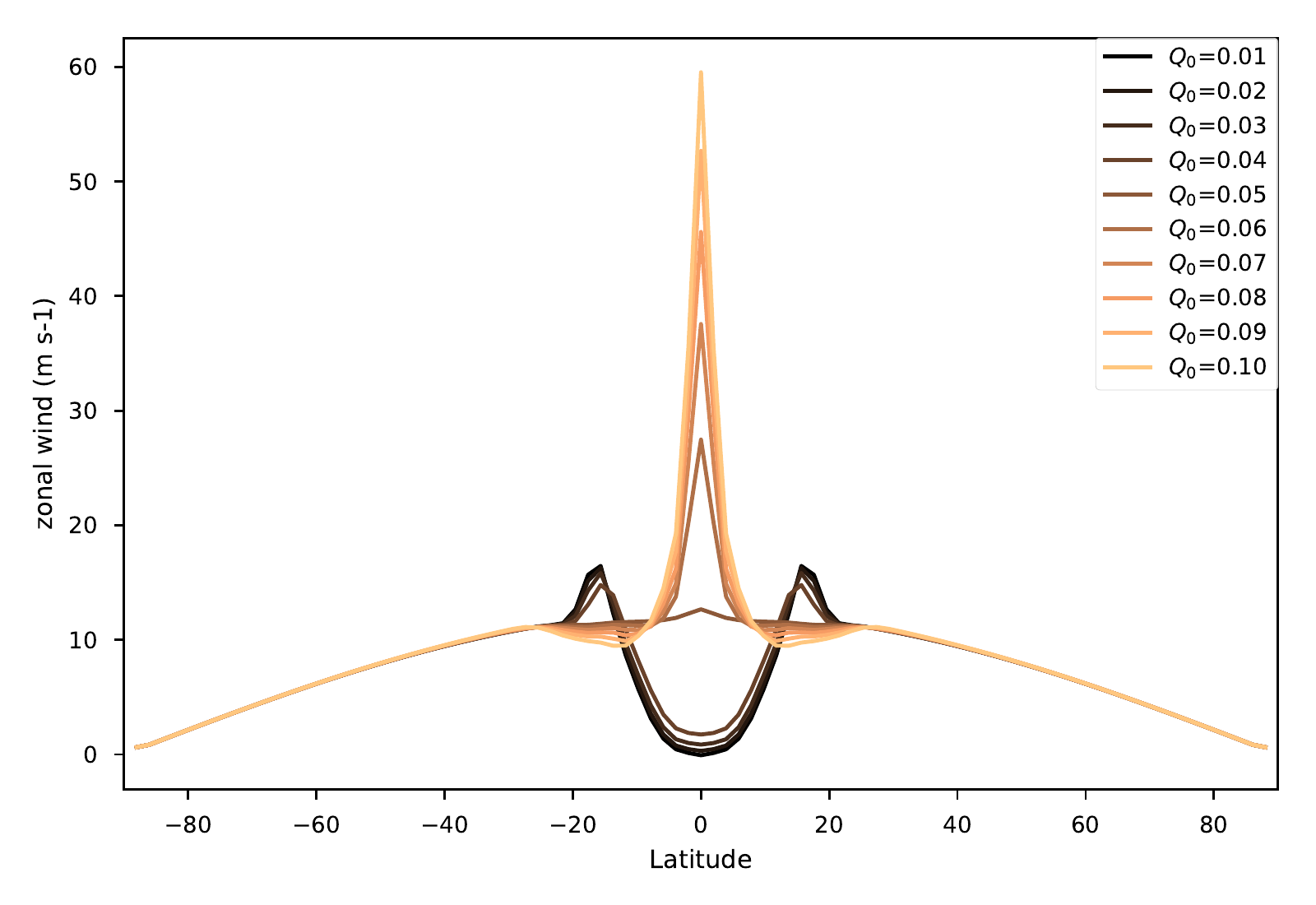}
  \includegraphics[width=0.45\linewidth]{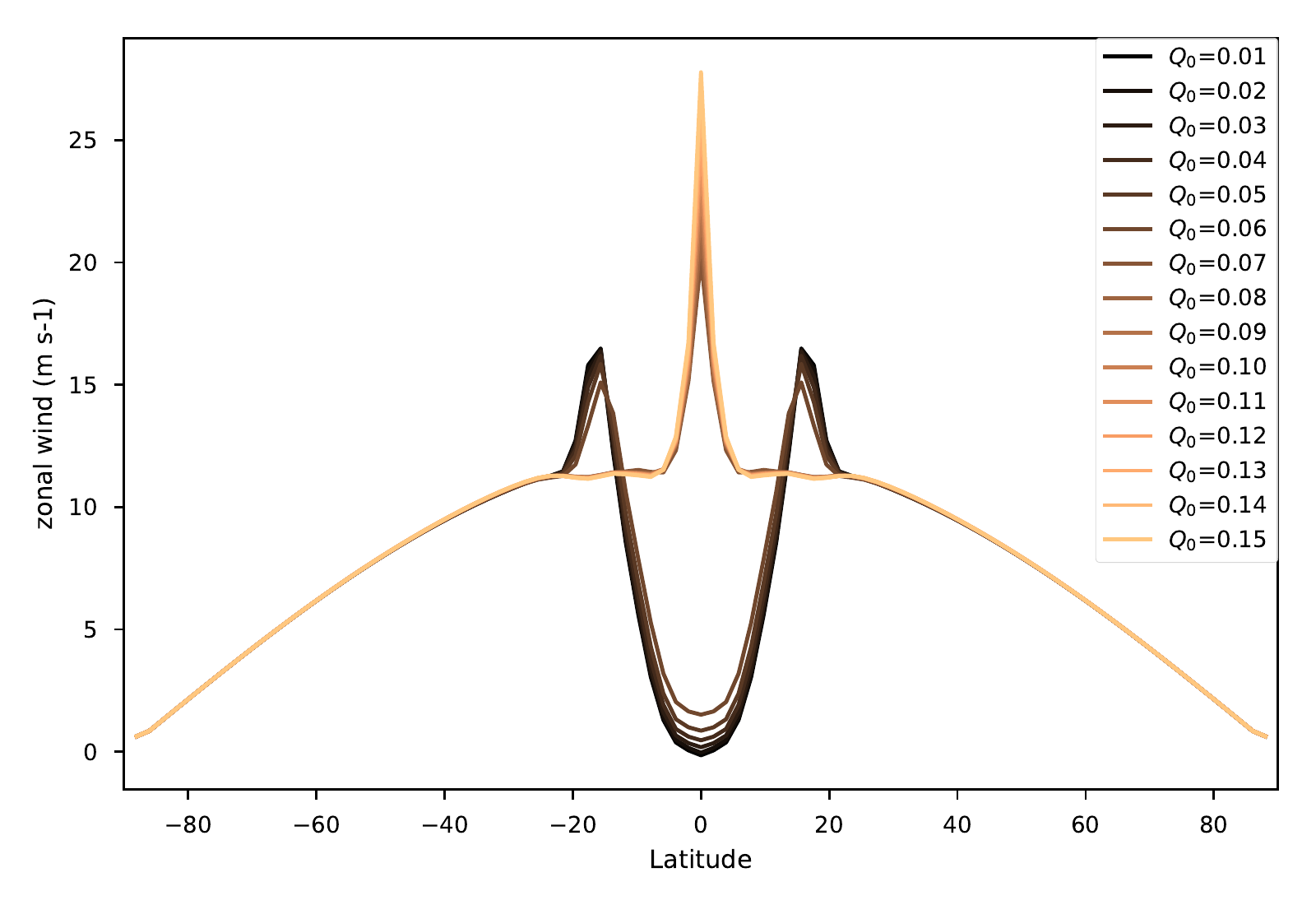}
  \caption{\label{fig:zonalwindpfl_hadley2} Vertically averaged zonal wind profile (2 vertical levels) for a constant eddy forcing (left) and resonant eddy forcing ($\epsilon=0.1$, right) for different forcing amplitudes.}
\end{figure*}
For instance, we show in Fig.~\ref{fig:zonalwindpfl_hadley2} the meridional profile of vertically averaged zonal wind computed with only 2 vertical levels, which is quite similar to the one depicted above with full vertical resolution (see Fig.~\ref{fig:zonalwindpfl_resonance} and Fig.~\ref{fig:zonalwindpfl_hadley}).
The main differences are that the transition to the superrotating regime seems even sharper and that stronger equatorial jets are obtained with 45 vertical levels.


Our experiments indicate that the hysteresis loop is quite sensitive to the vertical resolution, and depends on it in a non-monotonic manner.
This holds for both kinds of eddy forcings: resonant and constant.
Several hypotheses may be done to account for this sensitivity.
First of all, different choices of vertical levels result in different samplings of the vertical profile of the forcing.
Given the structure of the forcing, this amounts to multiplying the forcing amplitude by a constant factor.
Besides, since the Hadley cell feedback is proportional to vertical shear, poorly resolved vertical gradients may have large effects.
Finally, numerical modes of the discretized vertical diffusion operator may also play a part.

Ultimately, the hysteresis loop strongly depends on the number of vertical levels for low resolutions but converges for larger resolutions: for resolutions larger than 10 vertical levels (including a run with 90 levels), the hysteresis cycle does not change significantly.
In particular, all the hysteresis loops shown in the above sections have converged.

\subsection{Comparing the two types of bistability}\label{sec:comparison}

There are two major differences between bistability driven by the wave-jet resonance and by the Hadley cell: the behavior of the Hadley cell on the superrotating branch, and the sensitivity to vertical diffusion.

\begin{figure*}[ht]
  \centering
  \includegraphics[width=0.45\linewidth]{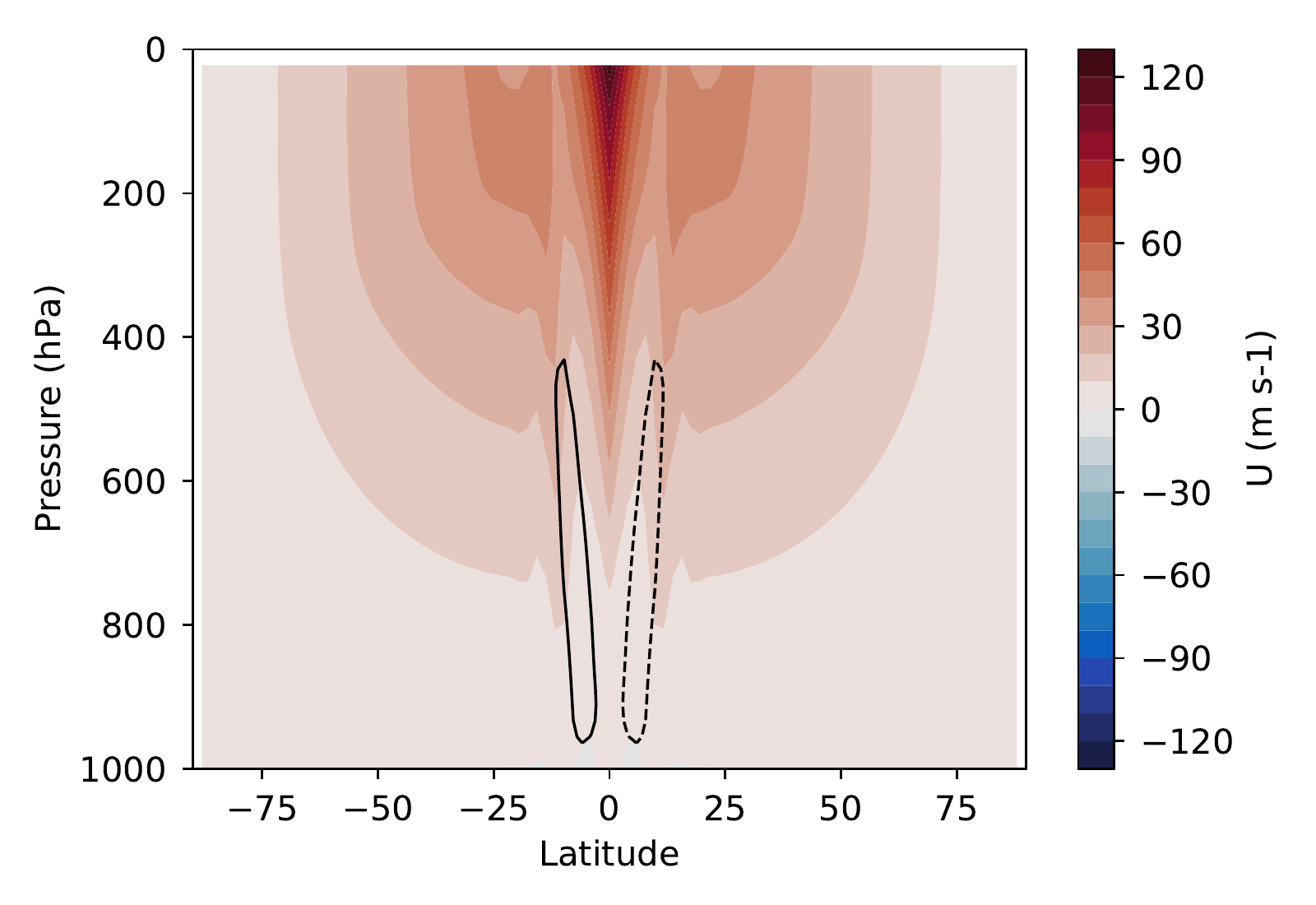}
  \includegraphics[width=0.45\linewidth]{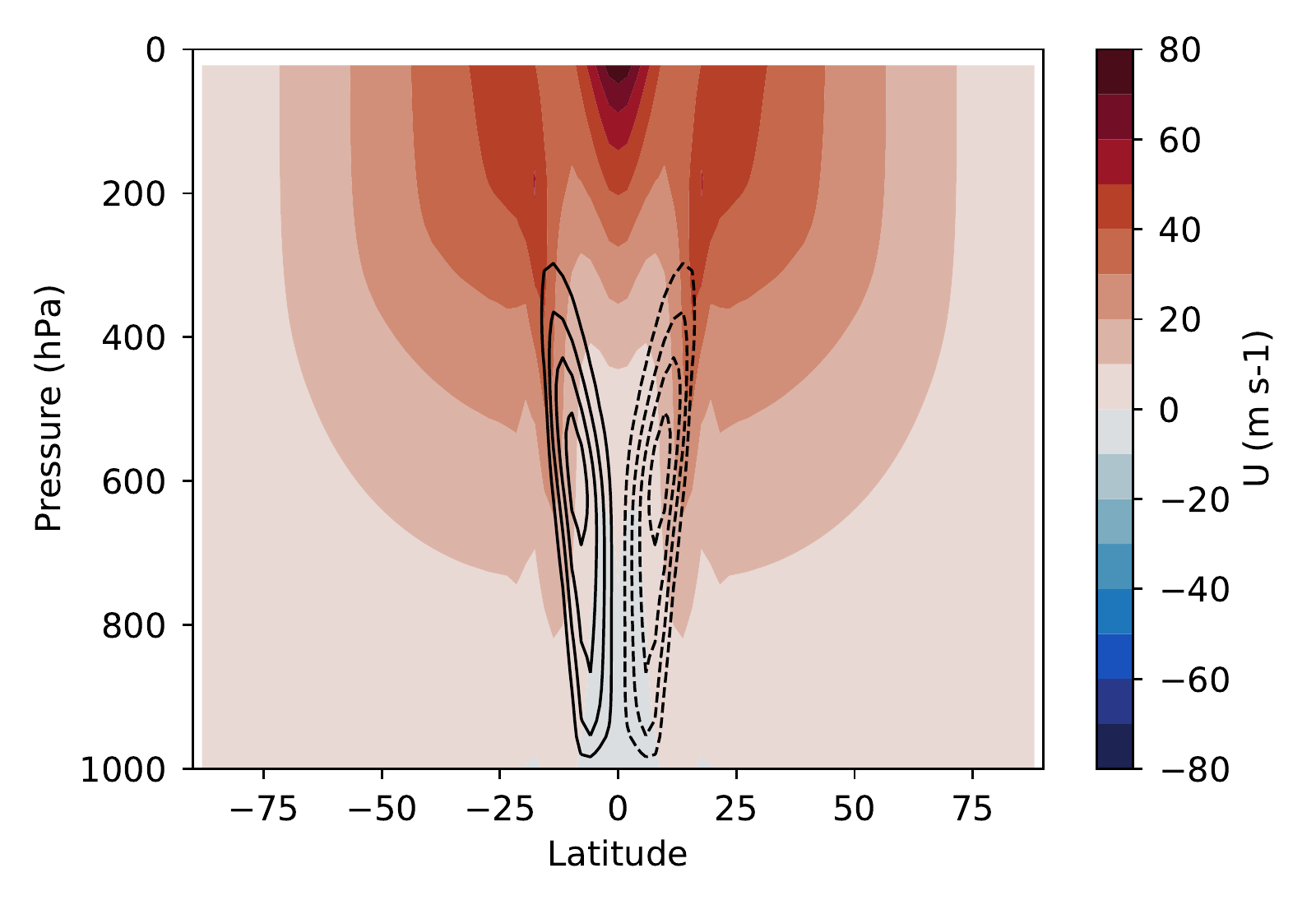}
  \caption{\label{fig:comparison} Comparison of the two types of superrotating states: Hadley-cell driven (left, collapsed Hadley cell) and resonance driven (right, Hadley cell not collapsed).}
\end{figure*}
Figure~\ref{fig:comparison} shows the 2D zonal wind field as well as the mean meridional circulation streamfunction for the two kinds of superrotating states: one on the upper-branch of the hysteresis loop obtained with a constant eddy forcing (with $Q_0 = 0.038$), and another on the upper-branch of the hysteresis loop obtained with a resonant eddy forcing ($\epsilon=0.1$, $Q_0=0.03$).
It illustrates the fact that the Hadley cell is almost as strong as in the conventional circulation in the resonance-induced superrotating state, while it is reduced by a factor 5 in the Hadley cell-induced superrotating state.
While in both cases, the equatorial jet is essentially confined to the upper troposphere, it is much sharper in the case of the constant forcing: both the vertical and the meridional wind shear are larger than in the resonant eddy forcing case.
The maximum velocity is also larger with the constant eddy forcing.
This is consistent with the behavior of the Hadley cell in the two cases.
If we further increase the resonant eddy forcing amplitude (not shown), we recover a state very similar to the superrotating state obtained with the constant eddy forcing at lower forcing amplitude, such as illustrated in the left panel of Fig.~\ref{fig:comparison}, with a sharper jet and collapsed Hadley cell.

Vertical momentum transport by the eddy viscosity is also expected to play an important role: we have seen in Sec.~\ref{sec:shell}\ref{sec:bistabilityhadley-balance} that, in the 0D model, when bistability is driven by the Hadley cell, it can be destroyed by increasing the strength of dissipative processes.
\begin{figure*}[ht]
  \centering
  \includegraphics[width=0.48\linewidth]{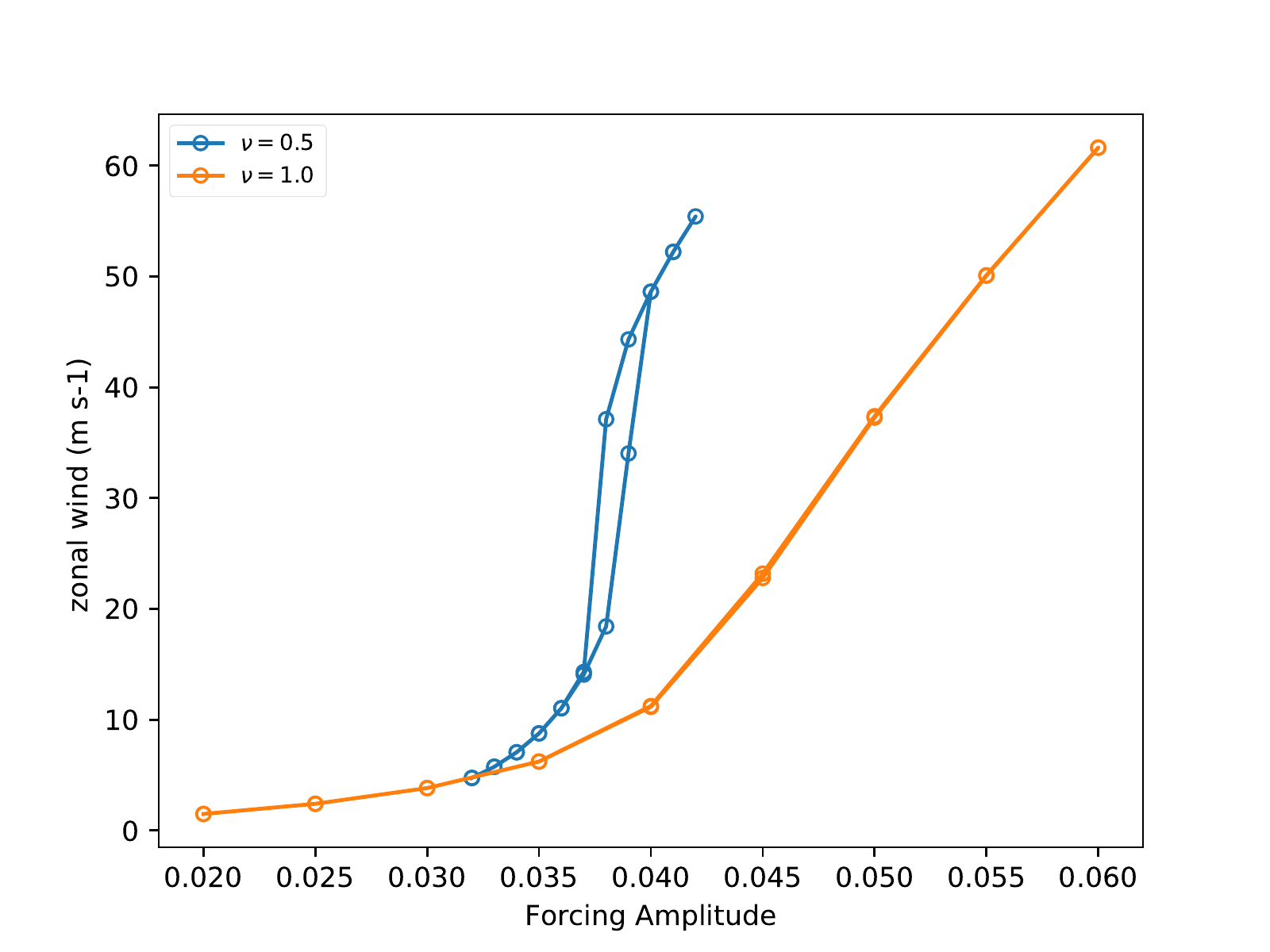}
  \includegraphics[width=0.48\linewidth]{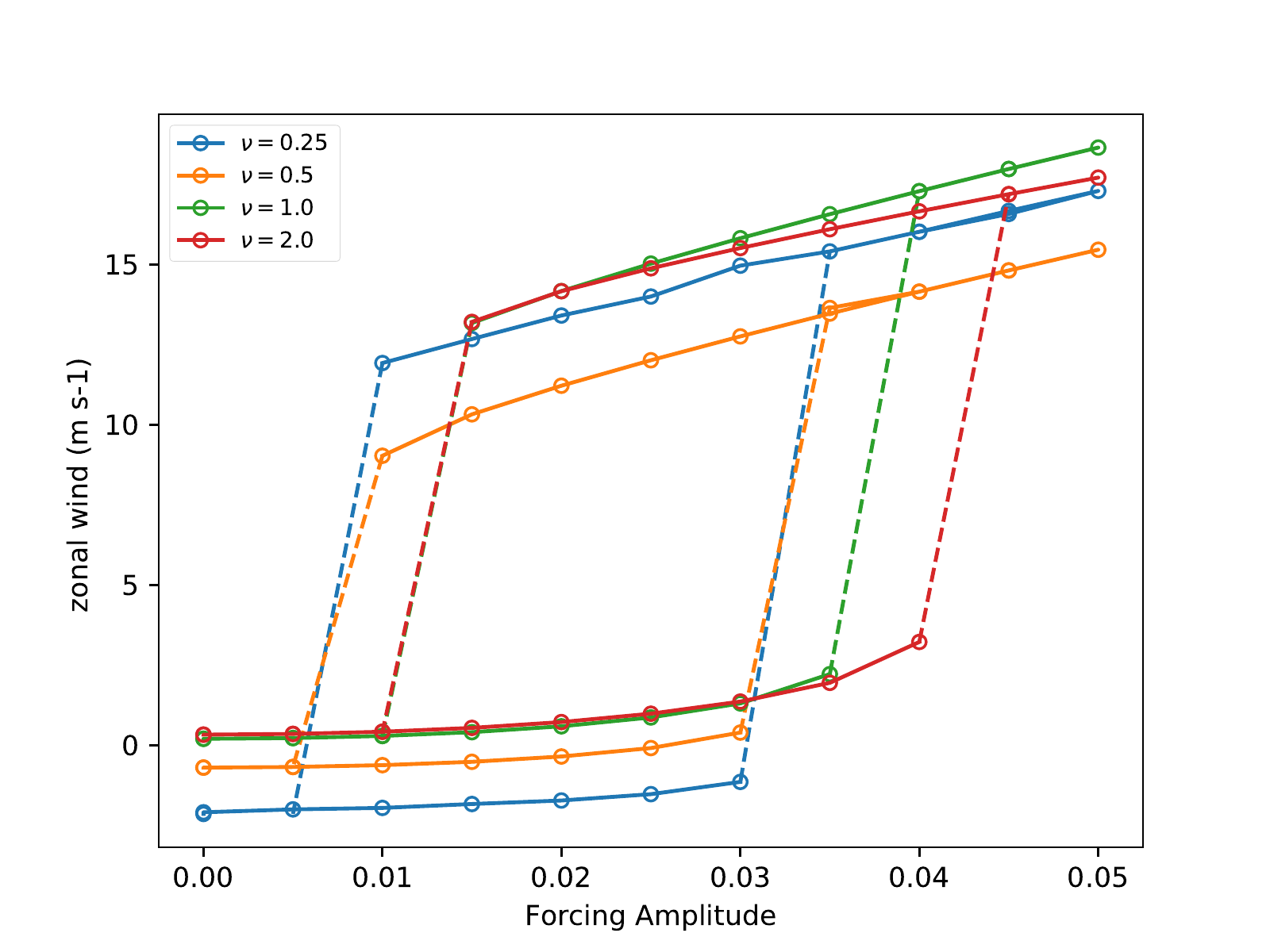}
  \caption{\label{fig:hysteresis_nu} Hysteresis experiments showing the effect of vertical momentum diffusion in the constant forcing case (left) and in the resonant forcing case (right, $\epsilon=0.1$).}
\end{figure*}
To test whether it is also the case in the 2D axisymmetric primitive equations, we show in Fig.~\ref{fig:hysteresis_nu} hysteresis experiments for several values of the vertical viscosity $\nu$, both for the Hadley-driven and the resonance-driven cases.
It is found that the Hadley-driven case exhibits high sensitivity of the stability thresholds of both the conventional and superrotating states (Fig.~\ref{fig:hysteresis_nu}, left).
We find that bistability disappears beyond a critical viscosity $\nu_c \approx 0.7$ m\textsuperscript{2}.s\textsuperscript{-1}.
On the other hand, bistability governed by the wave-jet resonance (Fig.~\ref{fig:hysteresis_nu}, right) is much more robust to variations of the vertical diffusivity than the Hadley-driven case: bistability subsists for vertical viscosities up to 2 m\textsuperscript{2}.s\textsuperscript{-1}, with an unaffected range of coexistence of the two states.
Again, this is in agreement with the theoretical analysis of Sec.~\ref{sec:shell}\ref{sec:bistabilitybalancequalitative}, where we have found that in the 0D model resonance-driven bistability subsists when friction is the main damping mechanism.

\section{Conclusion}\label{sec:conlusion}

In this paper, we have considered the question of atmospheric bistability at the planetary scale through the special case of equatorial superrotation.
This case is particularly interesting because it is frequently encountered in planetary atmospheres, and is hypothesized to have played a role in warm climates of the past on Earth.
From our point of view, a crucial point for the study of the transition to superrotation is the nature of the transition: continuous (akin to second-order phase transitions in condensed matter physics) or abrupt (first-order phase transition).
In the latter case, the transition may even occur spontaneously below the bifurcation point where the conventional state loses stability, driven by the fluctuations inherent to a turbulent atmosphere.
The mechanisms determining the nature of the transition may, but need not coincide with those maintaining the equatorial jet by converging angular momentum at the equator.
We have studied two such mechanisms corresponding to the two different cases.
On the one hand, it was suggested that an abrupt transition to superrotation may be triggered by a resonant response to non-zonal equatorial heating, which excites tropical waves which in turn accelerate the mean flow towards the east.
We have shown in an idealized model of zonal momentum balance at the equator that such a phenomenon indeed resulted in the appearance of multiple equilibria.
On the other hand, it was shown in an idealized framework that the Hadley cell itself could admit two different modes, which results in the coexistence of a conventional and a superrotating state when a constant torque is applied by an external operator.
In the same idealized framework, we have studied the interplay between the two mechanisms and showed that they differed by several characteristics: Hadley-driven bistability is relatively fragile, in the sense that it depends sensitively on vertical viscosity, while the resonance-driven bistability is much more robust to changes in this parameter.
On the other hand, the latter only occurs if the resonance is sufficiently peaked, which in the setup studied here amounts to sufficiently small linear friction.
Hence, the Hadley-driven bistability could be labelled as a large Ekman number, large Reynolds number regime, while resonance-driven bistability would correspond to a small Ekman number, arbitrary Reynolds number regime.
It should also be noted that, while most existing studies report a significant weakening of the Hadley cell in the superrotating state, our results indicate that in the resonance-driven case it is possible to obtain a superrotating state while maintaining a strong meridional circulation.
These findings are confirmed by numerical simulations of an axisymmetric primitive equations model, with an arbitrary number of vertical levels.
Based on these simulations, parameter values corresponding to the atmosphere of the Earth lie close to the boundary separating the two idealized regimes identified above.

These results may help shedding light on bistability and hysteresis (or the lack thereof) in full GCM simulations of superrotation.
Indeed, while abrupt transitions to superrotation have been reported before, it is often thought that such phenomena should be absent from state-of-the-art models.
Here, in the simpler case of a 2D axisymmetric model, we have observed unambiguously the existence of hysteresis phenomena.
We have also isolated some factors upon which bistability relies primarily: our simulations provide evidence for a very sensitive dependence on the vertical resolution, and on the damping mechanisms.
Further research is still needed to investigate whether the mechanisms described here hold in full 3D GCMs.
A critical factor differing from the framework considered here is that the eddy momentum convergence flux should depend dynamically on the full structure of the zonal wind field.
Here, we have considered a prescribed forcing, which, although consistent with diagnostics from 3D GCMs, was computed in a linear approximation assuming a uniform background wind, while there may be strong meridional shear in reality, especially in the superrotating state.
Understanding how that affects the results reported here would be key step towards providing a definitive answer to the question of the nature of the transition to superrotation.

\begin{acknowledgments}
  This project has received funding from the European Union's Horizon 2020 research and innovation programme under the Marie Sk\l odowska-Curie grant agreement No 753021.
  The research leading to these results has received funding from the European Research Council under the European Union's seventh Framework Programme (FP7/2007-2013 Grant Agreement No. 616811).
  Computer time was provided by PSMN.
\end{acknowledgments}

\bibliographystyle{ametsoc2014}
\bibliography{superrotation}

\end{document}